\documentclass[twocolumn]{aastex631}

\usepackage{lineno}

\newcommand{\kev}{\text{keV}}

\newcommand{\ergs}{\text{erg s$^{-1}$}}

\newcommand{\E}[1]{\times10^{#1}}

\newcommand{\U}[3]{$#1_{-#3}^{+#2}$}

\newcommand{\flux}{\text{erg cm$^{-2}$ s$^{-1}$}}

\newcommand{\fluxden}{$\mu$\text{Jy\,beam$^{-1}$}}

\shorttitle{Radio and X-ray observations of nearby central black holes}
\shortauthors{Urquhart et al.}

\graphicspath{{./}{figures/}}

\begin{document}

\title{X-ray and radio observations of central black holes in nearby low-mass early-type galaxies: Preliminary evidence for low Eddington fractions}

\author{Ryan Urquhart}
\affiliation{Center for Data Intensive and Time Domain Astronomy, Department of Physics and Astronomy,\\ Michigan State University, East Lansing, MI 48824, USA}
\author{Lauren I. McDermott}
\affiliation{Center for Data Intensive and Time Domain Astronomy, Department of Physics and Astronomy,\\ Michigan State University, East Lansing, MI 48824, USA}
\author{Jay Strader}
\affiliation{Center for Data Intensive and Time Domain Astronomy, Department of Physics and Astronomy,\\ Michigan State University, East Lansing, MI 48824, USA}
\author{Anil C. Seth}
\affiliation{Department of Physics and Astronomy, University of Utah, Salt Lake City, UT 84112, USA}
\author{Laura Chomiuk}
\affiliation{Center for Data Intensive and Time Domain Astronomy, Department of Physics and Astronomy,\\ Michigan State University, East Lansing, MI 48824, USA}
\author{Nadine Neumayer}
\affiliation{Max Planck Institut f\"ur  Astronomie, Heidelberg, Germany}
\author{Dieu D. Nguyen}
\affiliation{Universit\'e de Lyon, Universit\'e de Lyon1, Ens de Lyon, CNRS, Centre de Recherche Astrophysique de Lyon (CRAL) UMR5574, F-69230 SaintGenis-Laval, France}
\author{Evangelia Tremou}
\affiliation{National Radio Astronomy Observatory, Socorro, NM 87801 USA}


\begin{abstract}

We present new radio and X-ray observations of two nearby ($< 4$ Mpc) low-mass early-type galaxies with dynamically-confirmed central black holes: NGC\,5102 and NGC\,205. NGC\,5102 shows a weak nuclear X-ray source and has no core radio emission. However, for the first time we demonstrate that it shows luminous extended radio continuum emission in low-resolution, low-frequency ($< 3$ GHz) data, consistent with jet lobes on scales $\gtrsim 100$ pc formed from past accretion and jet activity. By contrast, in new, extremely deep, strictly-simultaneous Very Large Array and \textit{Chandra} observations, no radio or X-ray emission is detected from the black hole in NGC\,205. We consider these measurements and upper limits in the context of the few other low-mass early-type galaxies with dynamically-confirmed black holes, and show that the mean ratio of bolometric to Eddington luminosity in this sample is only $\textrm{log} \, (L_{\rm bol}/L_{\rm Edd}) = -6.57\pm0.50$. These Eddington ratios are \emph{lower} than typical in a comparison sample of more massive early-type galaxies, though this conclusion is quite tentative due to our small sample of low-mass galaxies and potential biases in the comparison sample. This preliminary result is in mild tension with previous work using less sensitive observations of more distant galaxies, which predict higher X-ray luminosities than we observe for low-mass galaxies. If it is confirmed that central black holes in low-mass galaxies typically have low Eddington ratios, this presents a challenge to measuring the occupation fraction of central black holes with standard optical emission line, X-ray, or radio surveys.

\end{abstract}

\keywords{}

\section{Introduction} \label{sec:intro}

The link between the evolution of host galaxies and the super-massive black holes (SMBHs) in their nuclei has been shown to be strong in massive galaxies (e.g., \citealt{2013ARA&A..51..511K}). The ubiquity of black holes in these environments has naturally led to the question of whether less massive galaxies also contain  central black holes. The question is not just an observational one of demographics, but bears on the broader question of how the massive galaxies acquired their SMBHs in the first place (e.g., \citealt{2010A&ARv..18..279V,2020ARA&A..58..257G,2020ARA&A..58...27I}). In some scenarios, the seed black holes that later grew to be SMBHs might be relatively common, which would be expected to lead to a high occupation fraction of central black holes in low-mass galaxies. If instead the initial SMBH seeds were formed less efficiently and hence were rarer, then present-day low-mass galaxies might have a lower occupation fraction.

It is clear that \emph{some} low-mass galaxies host central black holes, even down to the regime of ``intermediate-mass black holes" with $M \lesssim 10^5 \odot$ \citep{2020ARA&A..58..257G}, as revealed by optical (e.g., \citealt{2013ApJ...775..116R,2018ApJ...868..152B,2018ApJ...863....1C}), X-ray (e.g., \citealt{2015ApJ...799...98M,2018MNRAS.478.2576M,2021ApJ...922L..40L}), or radio-selected (e.g., \citealt{2020ApJ...888...36R}) active galactic nuclei (AGN). All of these studies individually find only that a small fraction ($\lesssim$1\%) of low-mass galaxies host observable AGN, though since the samples only partially overlap, the total fraction of observable AGN is higher.

The detected objects are typically rapidly accreting ($\gtrsim$1\% of their Eddington rate), and thus these studies are missing the likely much more numerous AGN in low-mass galaxies with lower accretion rates.  In nearby massive galaxies \citet{2009ApJ...699..626H} finds only a few percent accrete at above 1\% of their Eddington rate, while the median galaxy accretes at just 10$^{-5}$ of their Eddington rate.  In low mass galaxies, a low accretion rate makes it challenging to detect AGN. For example, a $10^5 M_{\odot}$ central black hole accreting at $\lesssim 10^{-5}$ of the Eddington luminosity has an X-ray luminosity $L_X \sim 10^{38}$ erg s$^{-1}$ and can be difficult or impossible to distinguish from X-ray binaries, supernova remnants, or other potential interloping sources from X-ray data alone.

An alternative approach is to perform comprehensive dynamical searches for central black holes in a volume-limited sample of galaxies. While conceptually appealing, the implementation of such a search is remarkably difficult: very fine spatial and spectral resolution is required to study the small spheres of influence of central black holes in low-mass galaxies, and only a modest number of such galaxies are within the reach of present instrumentation \citep{2021ApJ...921....8P}. This challenging approach has met with some success, dynamically confirming central black holes in the NGC\,5128 satellites NGC\,5102 and NGC\,5206 \citep{2018ApJ...858..118N}, the nearby field galaxy NGC\,404 \citep{2010ApJ...714..713S,2017ApJ...836..237N,2020MNRAS.496.4061D} and the M\,31 satellite NGC\,205 \citep{2019ApJ...872..104N}. These five galaxies represent a complete sample of the nearest early-type galaxies with stellar masses between 10$^{9-10}$~M$_\odot$, and suggest a high black hole occupation fraction in galaxies down to 10$^9$~M$_\odot$ \citep{2020ARA&A..58..257G}.

Since obtaining such dynamical measurements for a large sample of low-mass galaxies is currently infeasible, we turn back to multi-wavelength observations to explore whether a combination of data at different wavelengths---here, X-ray and radio observations---can be used to constrain the presence of central black holes accreting at low Eddington fractions in low-mass galaxies. We present new observations of two of the nearest low-mass galaxies with dynamically confirmed central black holes: NGC\,5102 (distance of 3.74\,Mpc; \citealt{2015ApJ...802L..25T}, and black hole mass of $\sim 9 \times 10^5 M_{\odot}$; \citealt{2019ApJ...872..104N}) and NGC\,205 (distance of 0.81\,Mpc; \citealt{2013AJ....146...86T}, and black hole mass of $\sim 6800 M_{\odot}$ with a substantial uncertainty; \citealt{2019ApJ...872..104N}). For NGC\,5102, we report new high resolution Karl G. Jansky Very Large Array (VLA) observations, paired with archival lower resolution Australia Telescope Compact Array (ATCA) and Australia Square Kilometer Array Pathfinder (ASKAP) images and archival {\it Chandra} data. For NGC\,205, we report new simultaneous deep VLA and {\it Chandra} X-ray observations. We combine the results from this analysis with published multi-wavelength data for several other dynamically-confirmed low-mass central black holes to assess the luminosity distribution of this class of source compared to other known classes of AGN.

\section{Data Analysis} \label{sec:data_analysis}

\subsection{NGC~5102}

\subsubsection{Radio}

We searched the Australia Telescope Online Archive (ATOA) for observations of NGC\,5102, and found that it had been observed on four occasions by the ATCA between November 2012 and February 2013 (project C2623, PI: S. Beaulieu). We analyzed these observations, each with an on-source time of $\sim$10\,h. All observations were taken with the Compact Array Broadband Backend (CABB; \citealt{2011MNRAS.416..832W}) correlator using a 2048\,MHz wide band centered at 2\,GHz (16\,cm band). PKS 1934$-$638 was used as the flux/bandpass calibrator, while PKS 1353$-$341 was used as the phase calibrator. 

Two of the observations were taken while the ATCA was in its extended 6A or 6B configuration, while the remaining two observations were taken in the more compact 1.5C or 1.5D configuration. We found that the 1.5C configuration (taken on 21 Nov 2012) observation was too noisy to be of use and thus do not include it in our final analysis.

Radio frequency interference (RFI) flagging and gain and phase calibration were performed using the Common Astronomy Software Application (CASA; \citealt{2007ASPC..376..127M}), version 5.4.1. We used the {\sc tclean} algorithm within CASA to image the data using a robust value of 1. Additionally, self-calibration was performed on the 1.5D observation from 20 Dec 2012. Due to the self-calibration process, a slight astrometric correction ($2-3\arcsec$) was required to re-align the 1.5D observations with the remaining ATCA images. Finally, to measure the spectral indices, all ATCA observations were split into two $\sim1\,$GHz-wide bands centered at 1.6 and 2.6\,GHz. 

Radio emission coincident with the center of NGC\,5102 is detected in all three of the ATCA observations, where we take the 
center of the nucleus to be the Gaia EDR3 optical position of:  ICRS (R.A., Dec.) = (13:21:57.610, --36:37:48.39) \citep{2016A&A...595A...1G,2021A&A...649A...1G}.

\begin{figure*}
    \centering
    \includegraphics[width=0.4\textwidth]{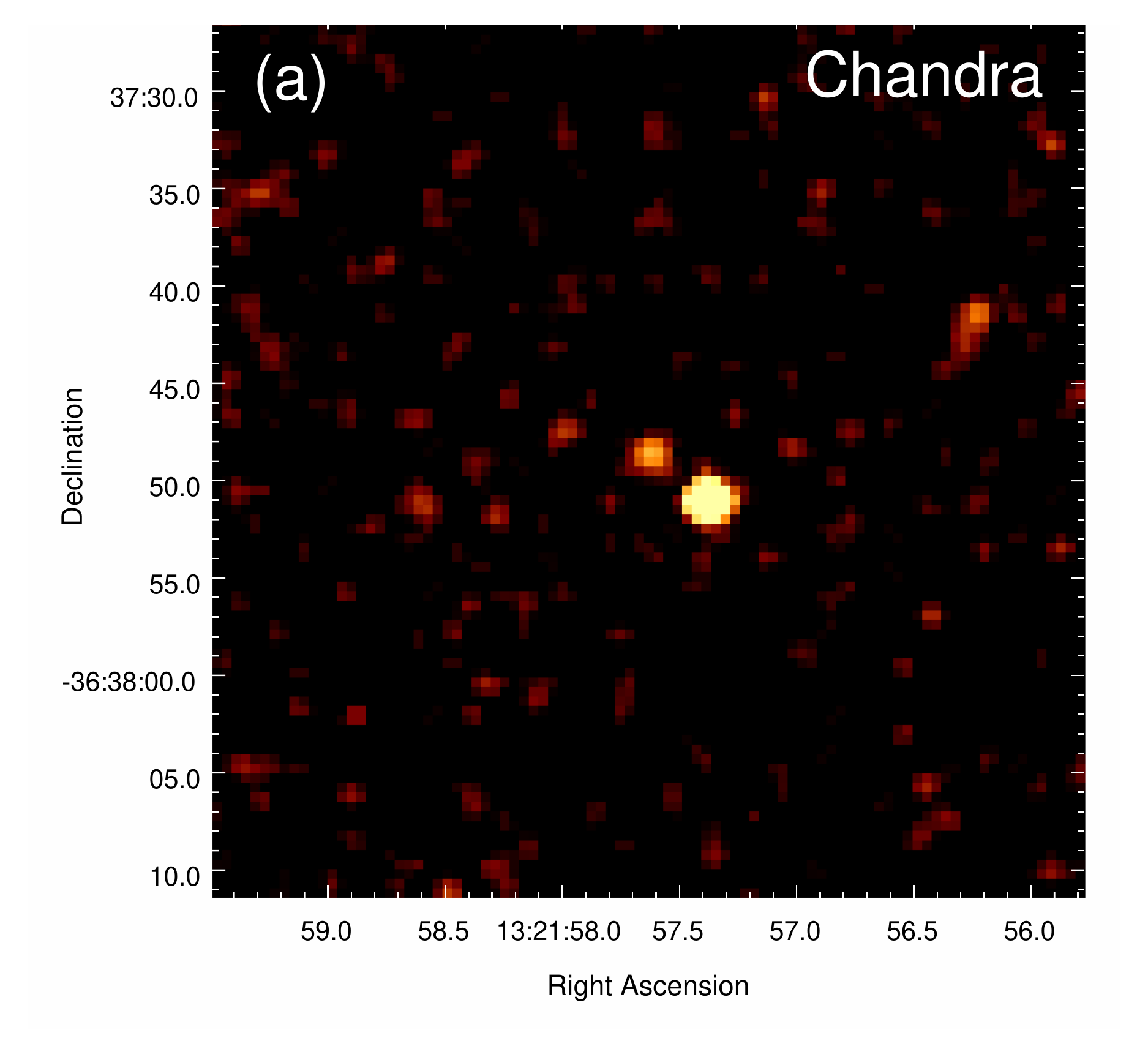}
    \includegraphics[width=0.4\textwidth]{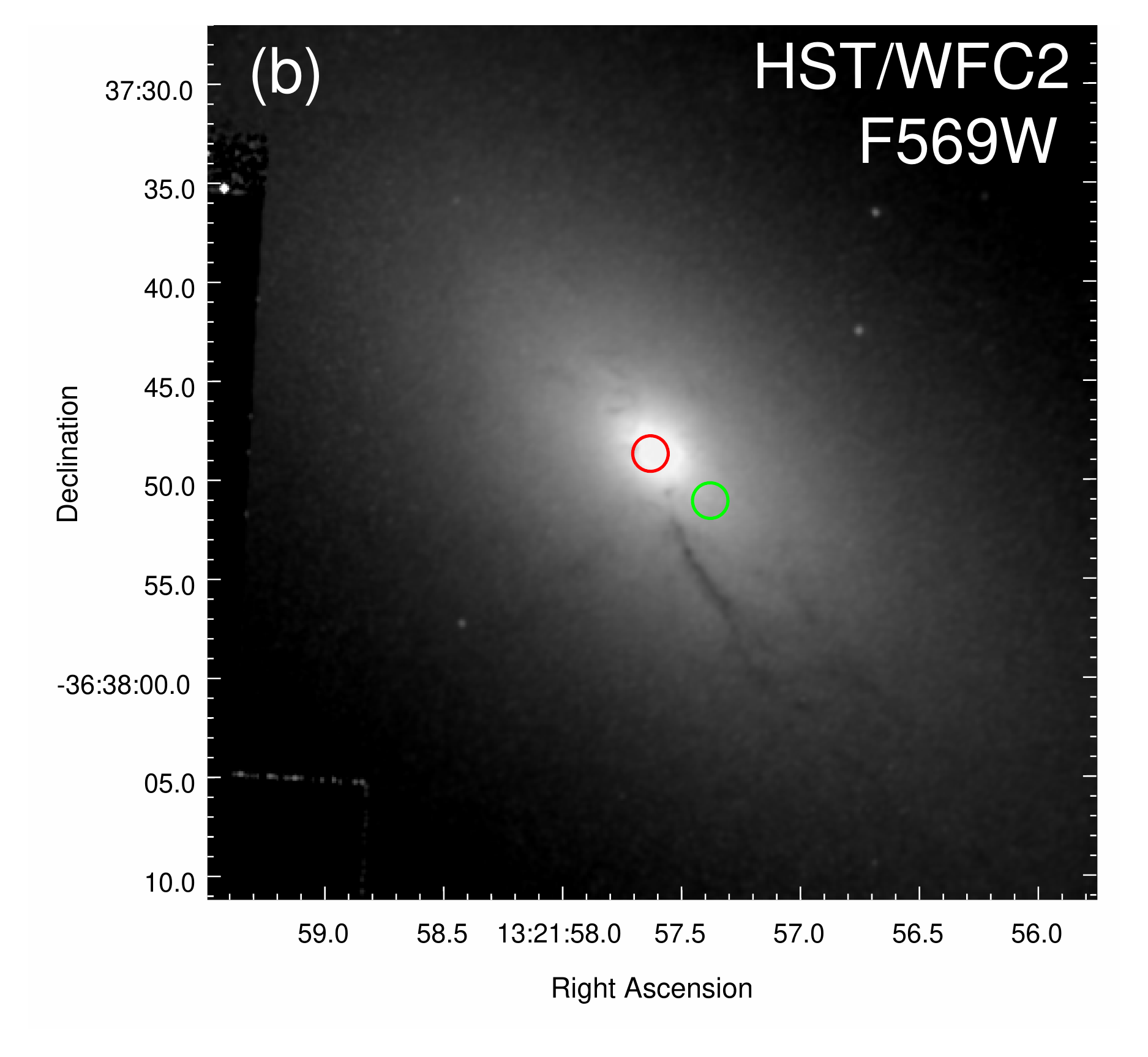}\\
    \includegraphics[width=0.3\textwidth]{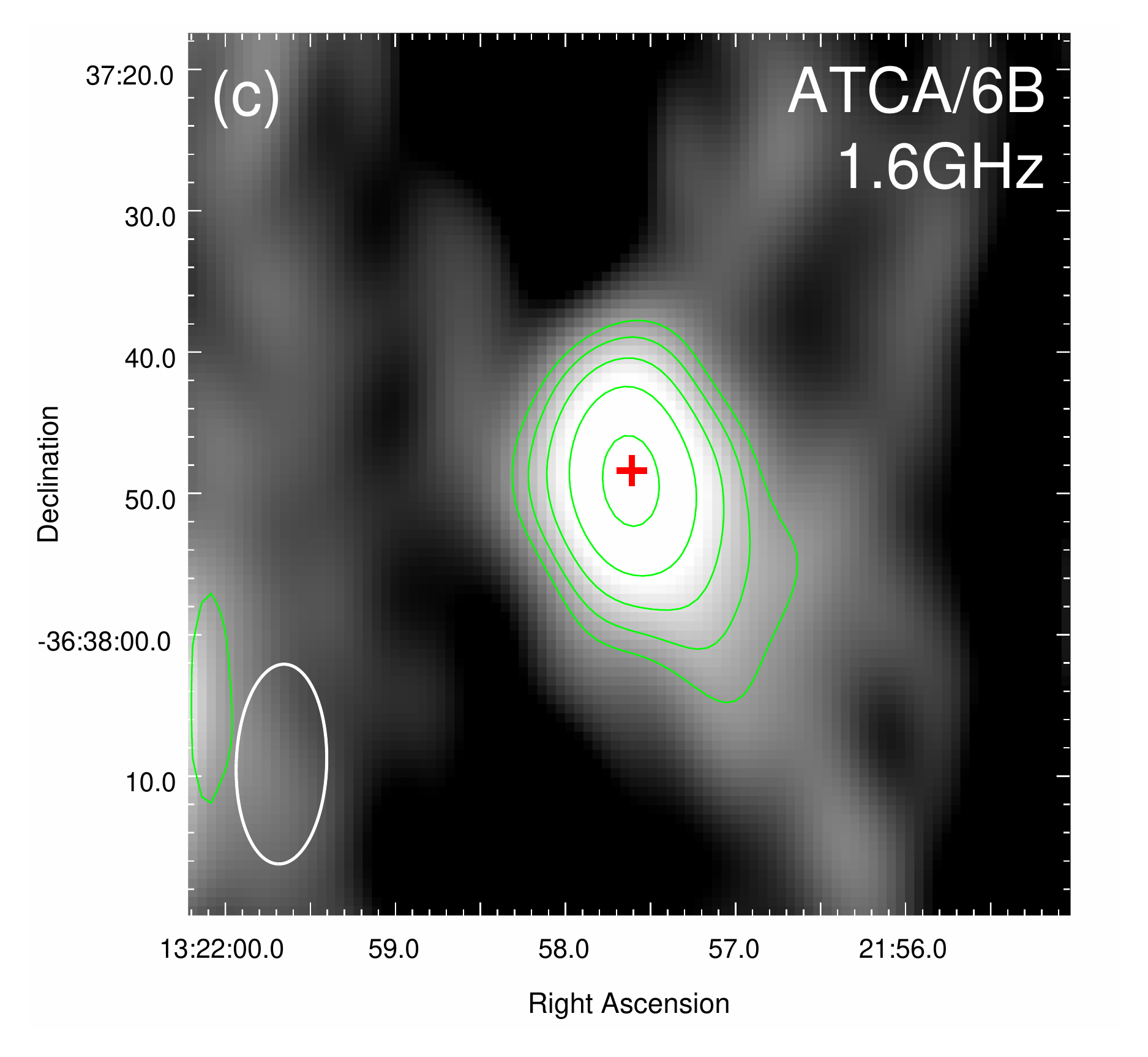}
    \includegraphics[width=0.3\textwidth]{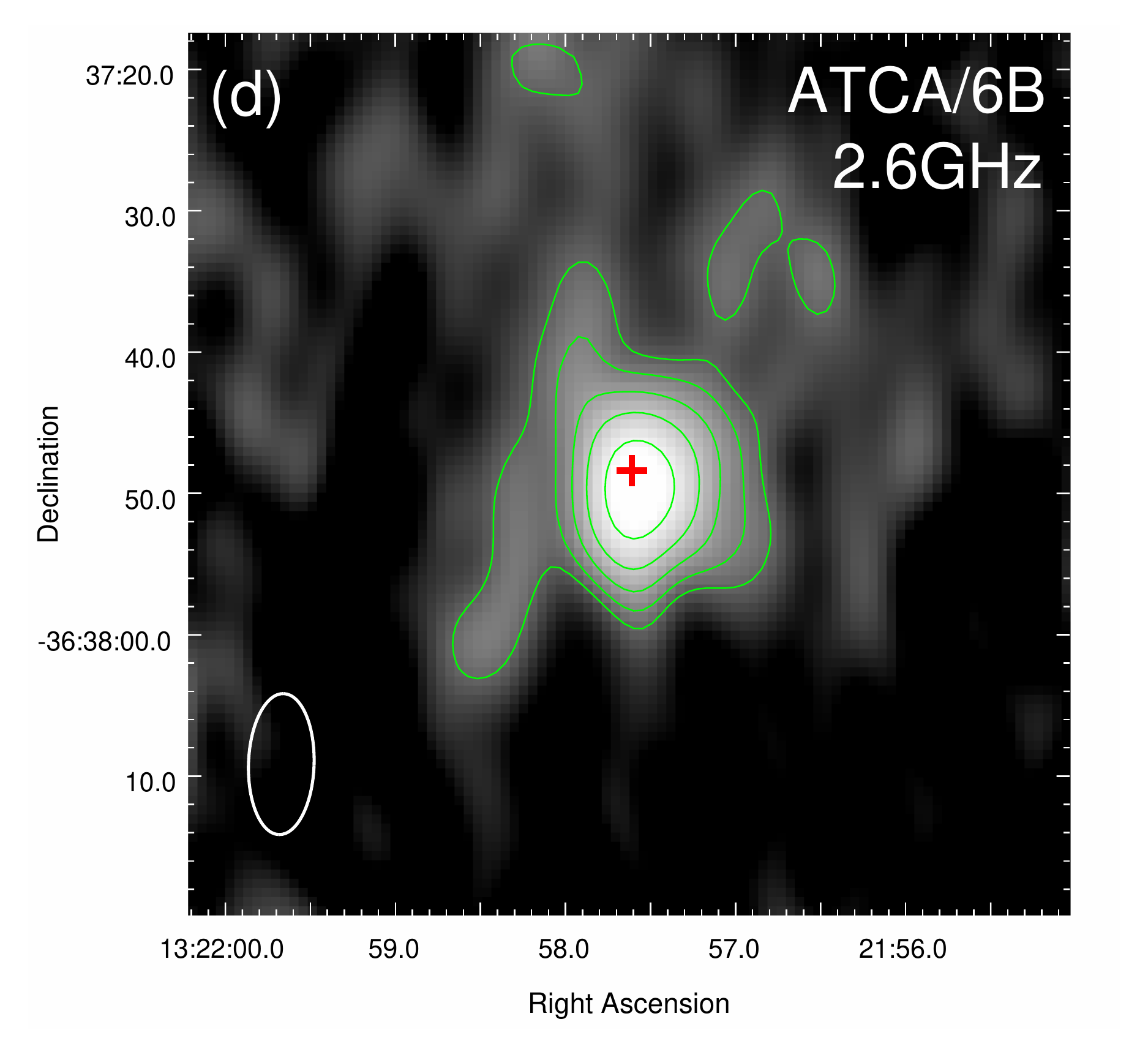} 
    \includegraphics[width=0.3\textwidth]{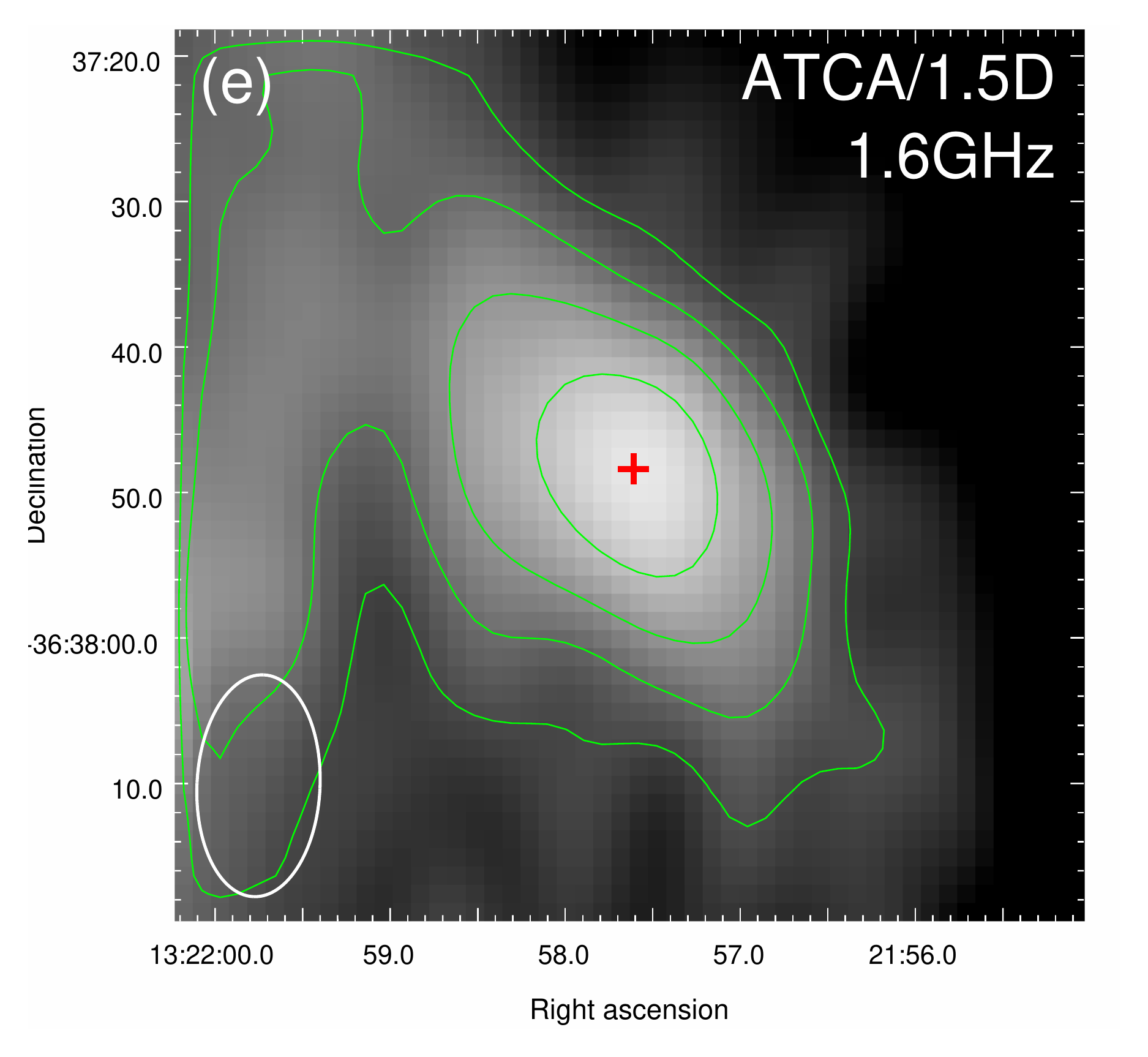}\\
    \includegraphics[width=0.3\textwidth]{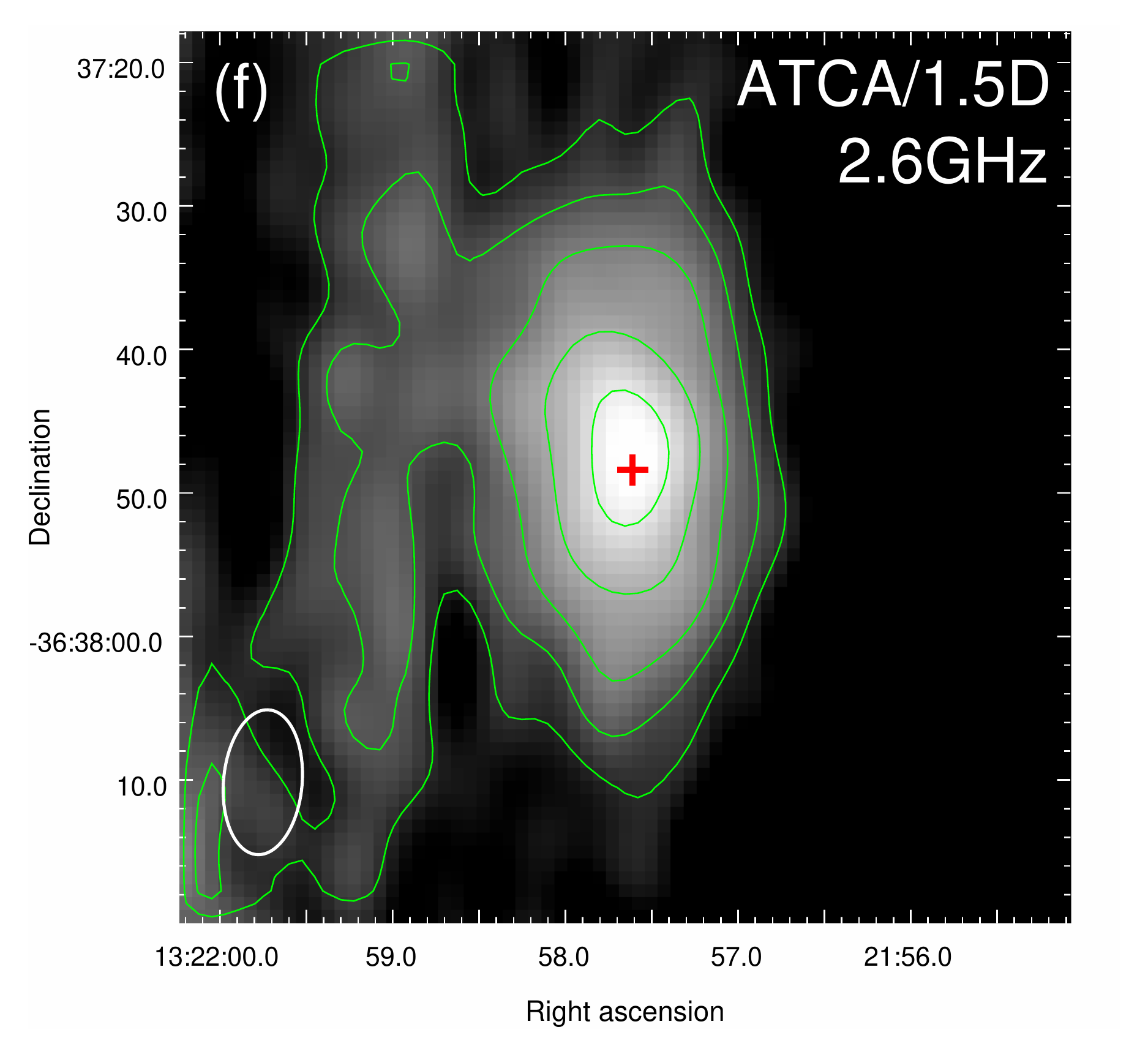}
    \includegraphics[width=0.3\textwidth]{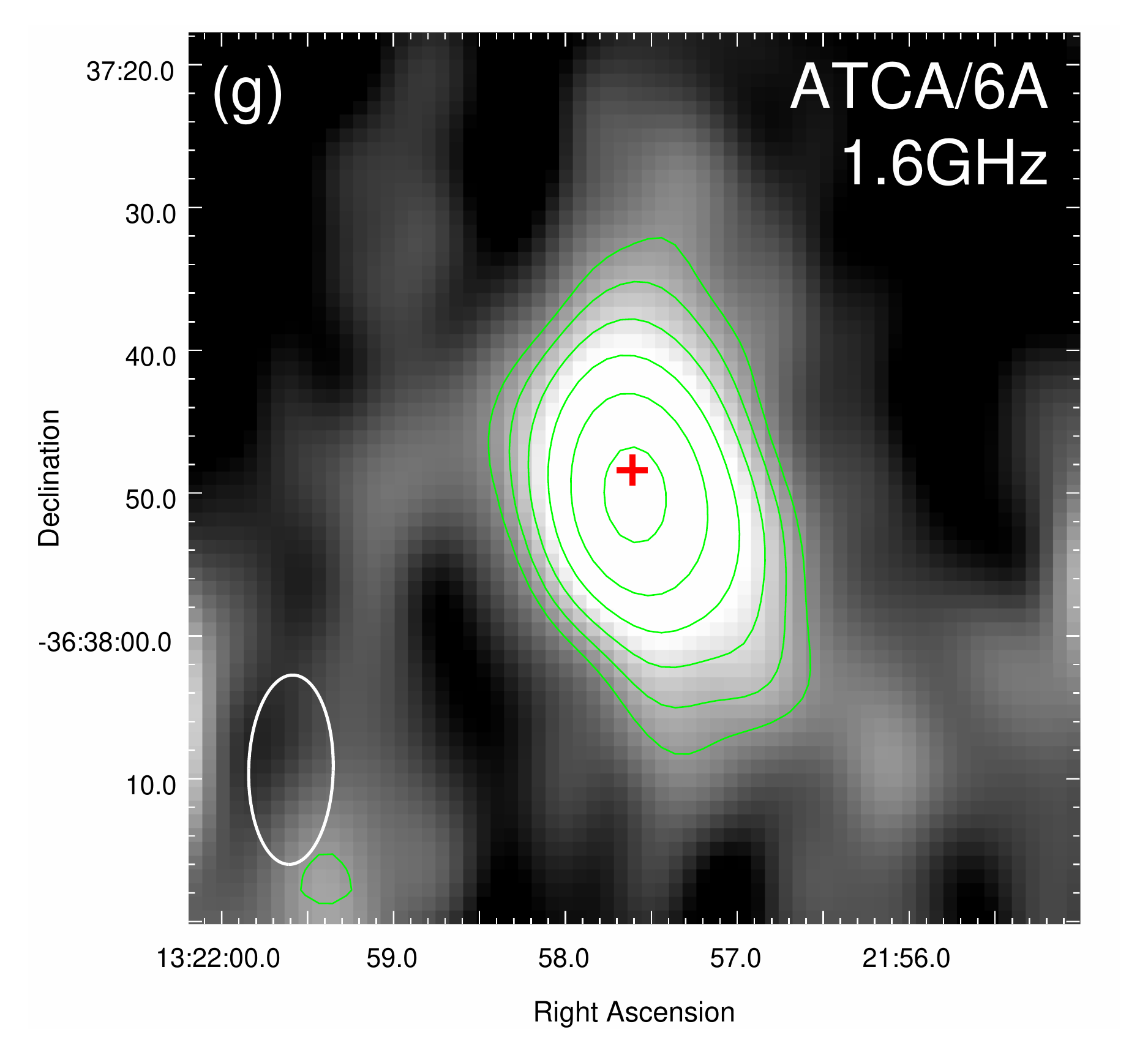}
    \includegraphics[width=0.3\textwidth]{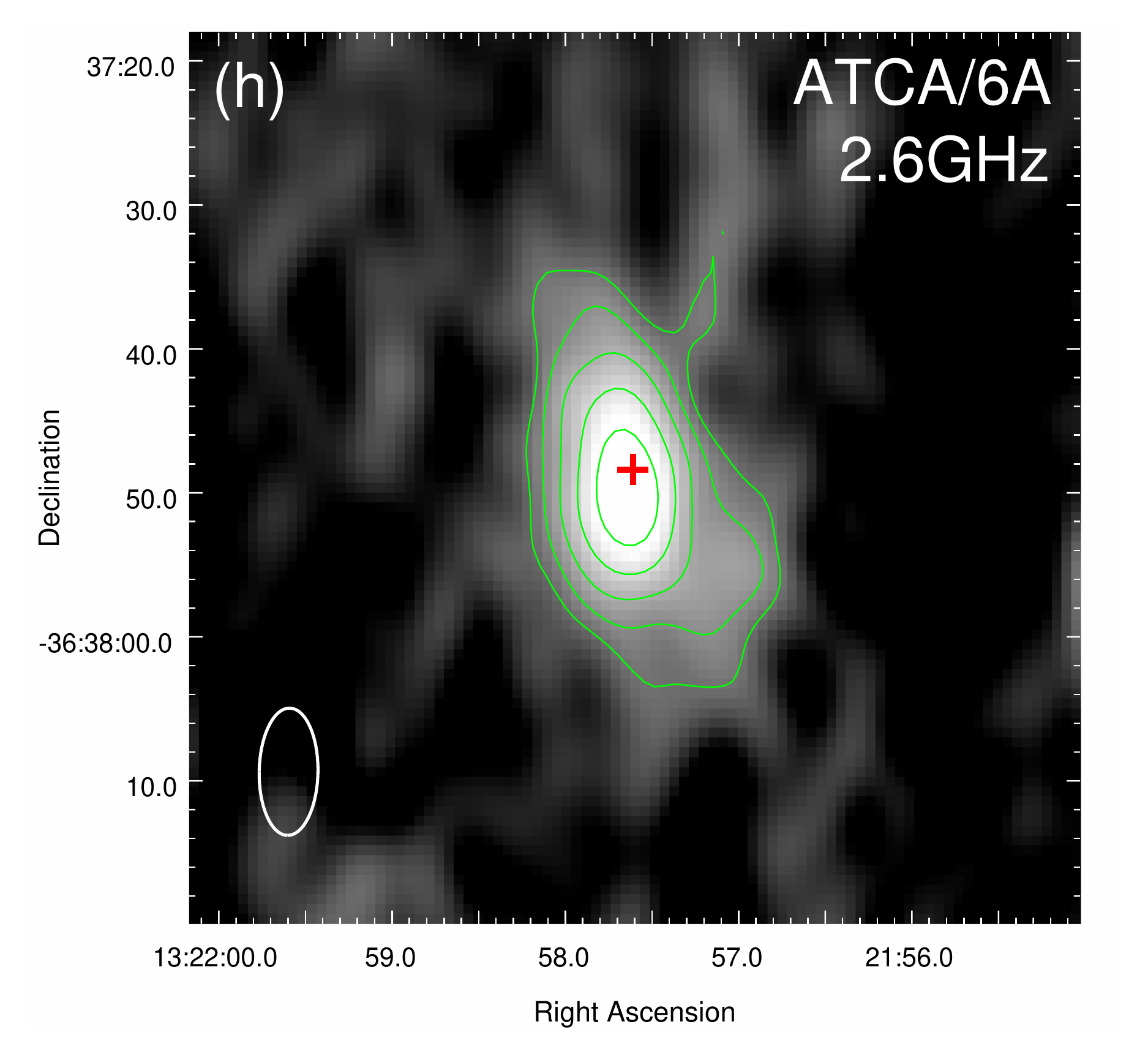}\\
    \includegraphics[width=0.3\textwidth]{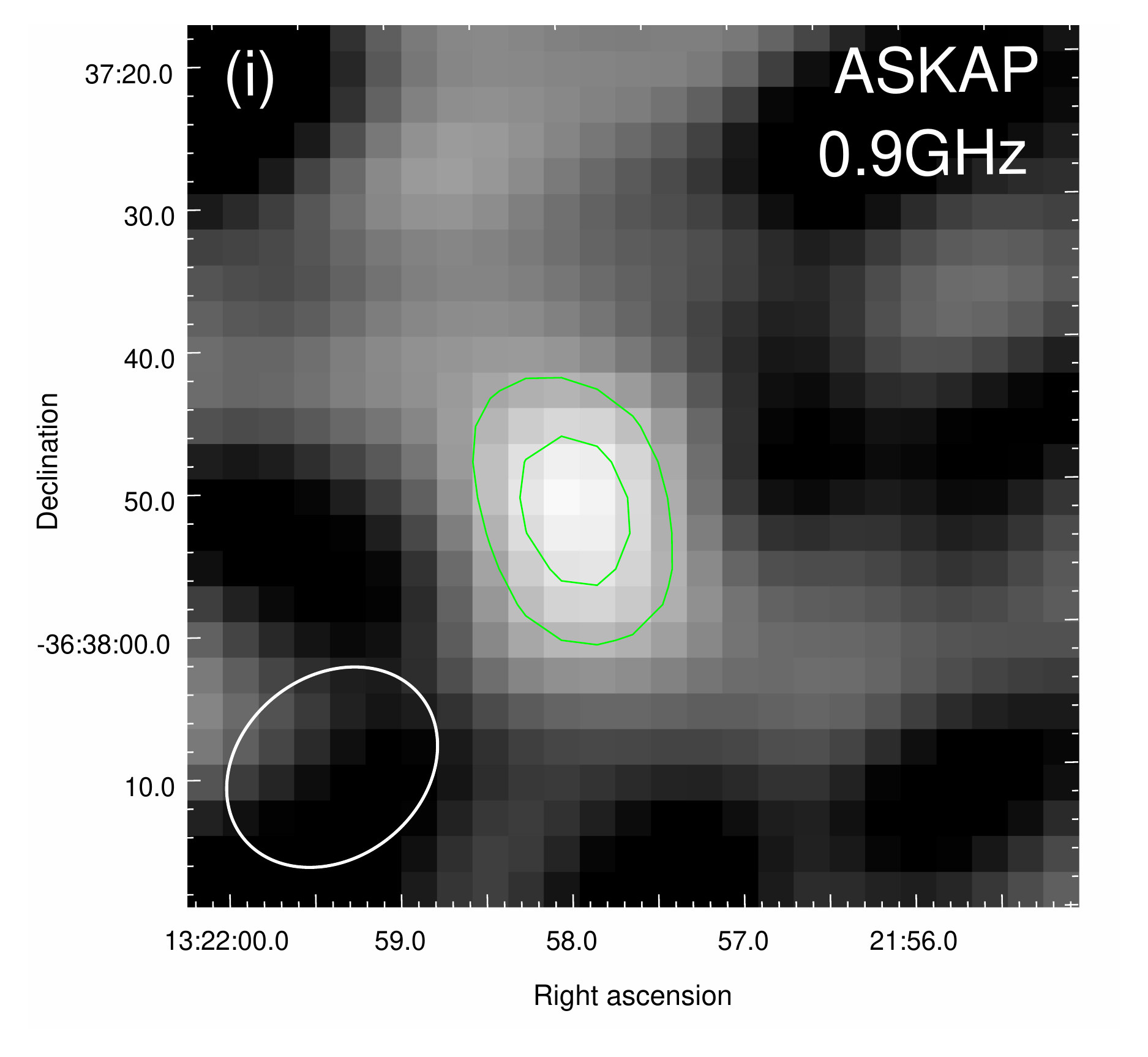}
     \includegraphics[width=0.3\textwidth]{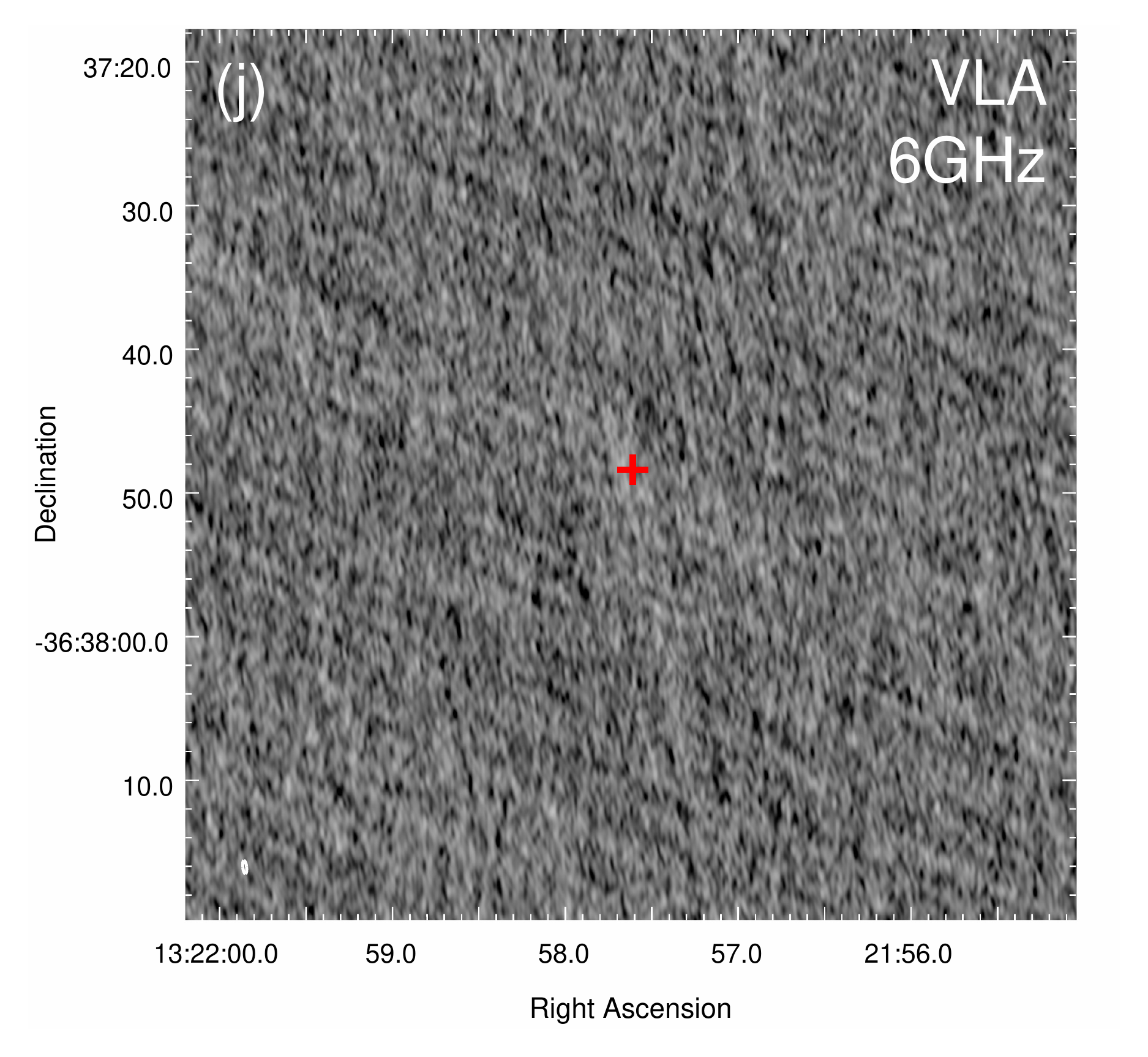}\\   
    \caption{The core of NGC\,5102, as imaged by \textbf{(a):} \textit{Chandra}/ACIS in the $0.3-7$\,keV band; \textbf{(b):} \textit{HST}/WFC2 in the F569W filter (the 0$\farcs$9 red and green error circles indicate the X-ray positions of the AGN and XRB, respectively); \textbf{(c):} ATCA in the 1.5D configuration at 1.6\,GHz. \textbf{(d):} ATCA in the 1.5D configuration at 2.6\,GHz. \textbf{(e):} ATCA in the 6A configuration at 1.6\,GHz. \textbf{(f):} ATCA in the 6A configuration at 2.5\,GHz. \textbf{(g):} ATCA in the 6B configuration at 1.6\,GHz. \textbf{(h):} ATCA in the 6B configuration at 2.6\,GHz. \textbf{(i):} ASKAP at 0.9\,GHz. \textbf{(j):} VLA at 6\,GHz. In the radio images, the red cross indicates the Gaia position of the nuclear source, and the FWHM extent of the synthesized beam is plotted as a white ellipse in the bottom left corner. The Gaia position is not included in panel (i) due to the uncertain astrometry. Green contours are set at $2\sqrt{2^n}\times \sigma $, where $\sigma$ is the local rms and $n=1,2,3,4..$.}
    \label{fig:n5102_images}
\end{figure*}

A summary of these observations, including the detected fluxes and restoring beams for each image, is given in Table \ref{tab:n5102_radio_obs}. Where the source is detected, and if the source is resolved, we estimate the size of the extended emission using 2D Gaussian fitting, deconvolved from the beam; the full-width at half-maximum (FWHM) is listed as Source Size in Table \ref{tab:n5102_radio_obs}. The ATCA images zoomed-in on the center of NGC\,5102 are shown in Figure \ref{fig:n5102_images}. At the distance of NGC\,5102 (0.81\,Mpc), 10\arcsec\ corresponds to about 180 pc.


We created a spectral index map using the low-resolution 1.5D configuration data, using the two sub-bands centered at 1.6\,GHz and 2.6\,GHz. First, we masked the two sub-band images to their respective 4$\sigma$ thresholds. Then, the 2.6\,GHz sub-band image was convolved to the restoring beam of the lower-resolution 1.6\,GHz sub-band image. We used the CASA tool {\sc imregrid} to re-grid the 1.6\,GHz image to match the pixel scale of the 2.6\,GHz image. Finally, we created a two-point spectral index map using the CASA task {\sc immath}. This image, centered on the core of NGC\,5102, is shown in Figure \ref{fig:n5102_spid}.

We followed up this detection with higher resolution observations of NGC\,5102, obtained with the VLA in its extended A configuration (project 20B-466, PI: J. Strader) on 27 Jan 2021. They consist of a short $\sim$1\,hr snapshot obtained with the C-band receiver (4$-$8\,GHz). Data were processed using the VLA CASA calibration pipeline, with additional manual flagging performed as necessary. 3C48 was used as the bandpass/flux calibrator, and J1316$-$3338 was used as the complex gain calibrator. The data were split into two 2.048\,GHz sub-bands, centered at 4.9\,GHz and 7.0\,GHz. No source was detected at the position of NGC\,5102 in these high-resolution observations (see Table \ref{tab:n5102_radio_obs} for details). Additionally, we imaged all 4$-$8\,GHz channels for a combined 4.096\,GHz-wide band, centered at 6\,GHz; even with the increased sensitivity due to the wider band no source was detected (Figure \ref{fig:n5102_images}, panel j).

Finally, we downloaded calibrated, publicly-available data from CSIRO's Australia Square Kilometer Array Pathfinder (ASKAP) archive\footnote{https://research.csiro.au/casda/}. One observation of NGC\,5102 was taken on 29 Apr 2019 as part of the Rapid ASKAP Continuum Survey (project/ID AS110/8593; \citealt{2020PASA...37...48M}). The observation had a duration of approximately 15 minutes and used a 288\,MHz wide band centered at 887.5\,MHz. A radio point source is detected coincident with the center of NGC\,5102 to within a few arcsec: the details of this radio emission are given in Table \ref{tab:n5102_radio_obs}. We note that there appear to be astrometric variations of up to $\sim 2-3\arcsec$ across the ASKAP field when comparing it to the higher-resolution ATCA and VLA images.
These variations are inconsistent even among nearby sources. The causes are not immediately clear, but might partially be due to frequency-dependent self-calibration errors, and indeed the first Rapid ASKAP Continuum Survey paper discusses larger-than-predicted variations in the survey astrometry 
\citep{2020PASA...37...48M}. For the purpose of this paper, we take the flux density measurement from the ASKAP data as reliable, but do not make any conclusions based upon its astrometry at the few arcsec level.


\subsubsection{X-ray}

NGC\,5102 was observed by \textit{Chandra} on 2002 May 21 for 34\,ks (ObsID 2949, PI: Kraft) with the ACIS-S detector. We reprocessed this observation using standard tasks with the Chandra Interactive Analysis of Observations ({\sc ciao}) software package v4.11 \citep{2006SPIE.6270E..1VF}. Any intervals of high particle background were filtered out.  Imaging analysis is performed with HEASARC's SAOImage {\sc ds9} v8.1 \citep{2003ASPC..295..489J}.

We detect two sources near the optical center of NGC\,5102 (Figure \ref{fig:n5102_images}, panel a). For each, source counts were extracted from a 1.48$^{\prime\prime}$ radius, corresponding to a 0.902 encircled energy fraction (for a point source). Local background counts were taken from source-free regions with a radius three times larger than the source extraction region. We measure a count rate of $(5.1^{+2.6}_{-2.0})\times10^{-4}$ ct s$^{-1}$ (errors represent 90\% confidence) for the fainter nuclear source and $(36\pm6)\times10^{-4}$ ct s$^{-1}$ for the brighter off-nuclear source, which is consistent with previous work \citep{2005ApJ...625..785K}. Due to the low number of source counts, robust spectral and timing analyses were not possible. In order to convert the count rate to an unabsorbed flux and luminosity, we assume a simple absorbed power-law (photon index $\Gamma = 1.7$; e.g., \citealt{1995ApJ...455..623C}; typical of sources in the low/hard accretion state), and use the Chandra X-Ray Center online installation of PIMMS\footnote{https://cxc.harvard.edu/toolkit/pimms.jsp} (Version 4.11a). In estimating the X-ray luminosity (Table \ref{tab:lrlx}), we use a line-of-sight column density $n_{H}=4\times10^{20}$ particles cm$^{-2}$ \citep{nh16} and distance $D=3.74\,$Mpc \citep{2015ApJ...802L..25T}. 

We use the {\sc ciao} task \textit{wavdetect} to find the positions of both X-ray sources. The nuclear source has an X-ray ICRS position of R.A.=13h21m57.628s, Dec.=$-36^{\circ}37^{\prime}48.68^{\prime\prime}$. The source to the south-west has a position R.A.=13h21m57.377s, Dec.=$-36^{\circ}37^{\prime}51.06^{\prime\prime}$. Positional errors are determined using Equation (5) from \citet{2005ApJ...635..907H}. This is combined, in quadrature, with the absolute instrumental astrometry of \textit{Chandra}/ACIS ($\approx0\farcs8$ at 90\% confidence\footnote{https://cxc.harvard.edu/cal/ASPECT/celmon/}) as there are insufficient point source coincidences to match the image with a standard astrometric frame. The resulting positional uncertainty for both sources is 0.8$^{\prime\prime}$. Both sources match, within errors, to previous work by \citet{2005ApJ...625..785K}; they are sources 28 (nuclear source) and 27 (south-west source) in their Table 2.

The nuclear X-ray source matches the high-precision optical position to within 0.4\arcsec, well within the $1\sigma$ uncertainty of the X-ray position, implying that this X-ray source is indeed associated with the nucleus.

\subsection{NGC~205}

\subsubsection{Radio}

We observed NGC\,205 as part of a joint VLA and \textit{Chandra} project. Two radio observations were taken on back-to-back days (2020 Oct 6 and 7) while the VLA was in its most extended A configuration (NRAO proposal code SL0277). Each observation was $\sim4.5$\,hr long and utilized the 3-bit samplers at C band, obtaining two 2.048 GHz wide basebands that covered the full 4--8 GHz range. Similar to NGC\,5102, we used data processed by the VLA CASA calibration pipeline. 3C48 was used as the bandpass/flux calibrator, and J0038+4137 was used as the phase calibrator. 

Since no significant radio continuum source was found near the center of NGC\,205, we chose to maximize the signal-to-noise by stacking the data in both time and frequency, resulting in a single image centered on 6\,GHz.  The stacked image reached a local rms of $1.1\,\mu$Jy\,beam$^{-1}$ with a 
1.45\arcsec$\times$1.08\arcsec\ FWHM Gaussian restoring beam, implying a $3\sigma$ upper limit of $< 3.3 \mu$Jy (the previous best limit, taken at 1.5\,GHz, was $<22.6\,\mu$Jy; \citealt{2019ApJ...872..104N}). We took the center of the optical nucleus to be that expected for emission from a black hole: an ICRS (R.A., Dec.) = (00h40m22.054s, $+41^{\circ}41^{\prime}07.50^{\prime\prime}$), which is a \emph{Gaia} DR2 measurement taken from \citet{2019ApJ...872..104N}.

\subsection{X-ray}

An $\sim 80$ ksec \textit{Chandra}/ACIS-S observation of NGC\,205 was taken to accompany the VLA observations. The observation was split into two $\sim40$\,ksec chunks obtained from 2020 Oct 6 to 8  (ObsIDs 22585 and 24754, respectively). The first of these observations, spanning 2020 Oct 6 to Oct 7, fully covered and hence was strictly simultaneous with the second of the two VLA blocks.

As before, data were reprocessed using standard tasks within {\sc ciao} and intervals of high background flaring were filtered out. We imaged both observations using HEASARC's {\sc ds9} visualization package. No X-ray source was detected in either image.  To maximize sensitivity, we merged both observations using the {\sc ciao} task \textit{merge\_obs}. This resulted in a total exposure time of 80\,ks, but still no X-ray source coincident with NGC\,205 was identified. Using the method of \citet{1991ApJ...374..344K}, we place a 90\% upper limit on the count rate of $<3.8\times10^{-5}$ count\,s$^{-1}$. We use PIMMS to estimate the flux upper limit, once again assuming a standard absorbed power-law with photon index $\Gamma=1.7$ and line-of-sight column density $n_H=6.2\times10^{20}$\,cm$^{-2}$ \citep{nh16}. We use a distance of $D=0.81$\,Mpc \citep{2013AJ....146...86T}.

\begin{figure*}[t]
    \centering
    \includegraphics[width=0.48\textwidth]{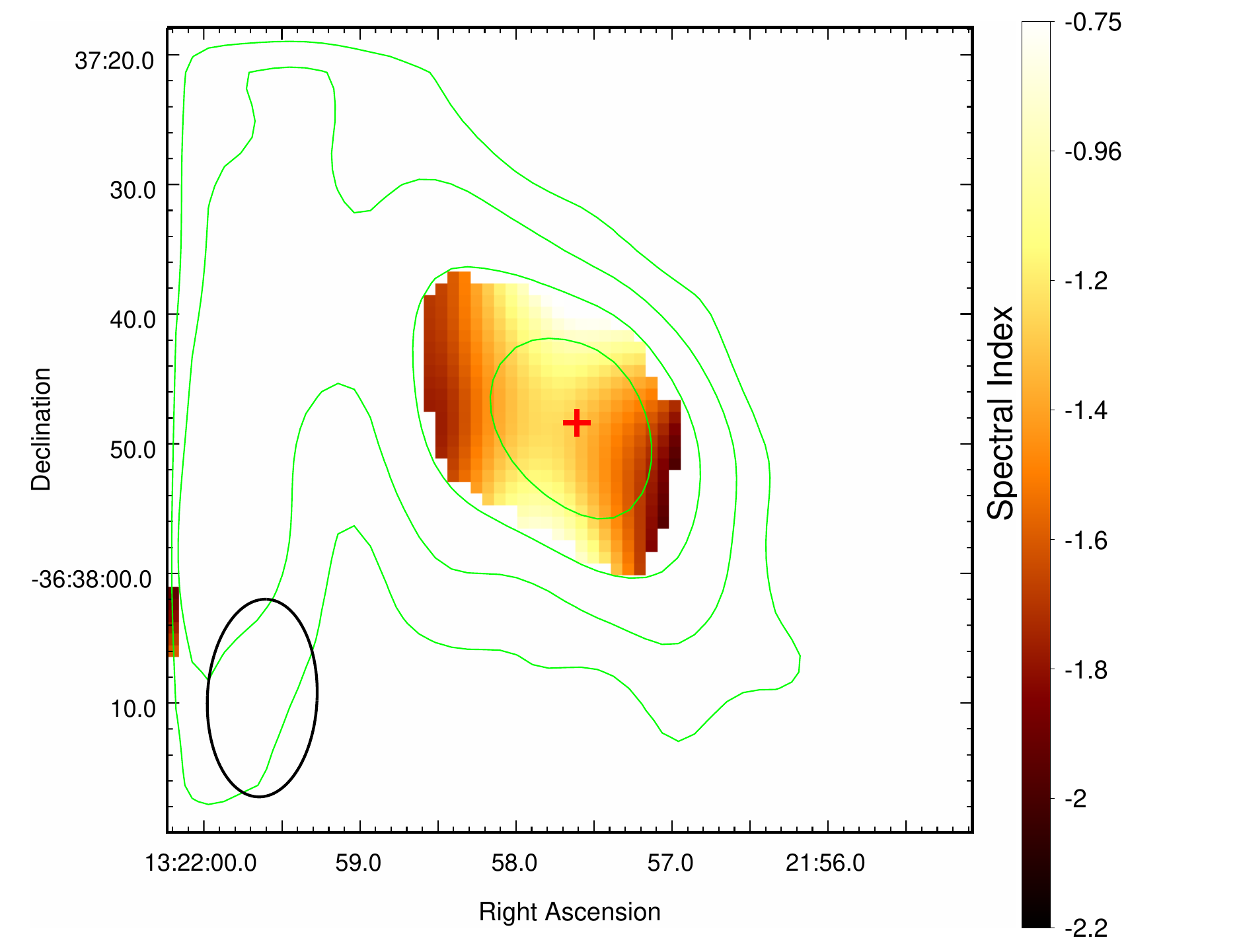}
    \includegraphics[width=0.48\textwidth]{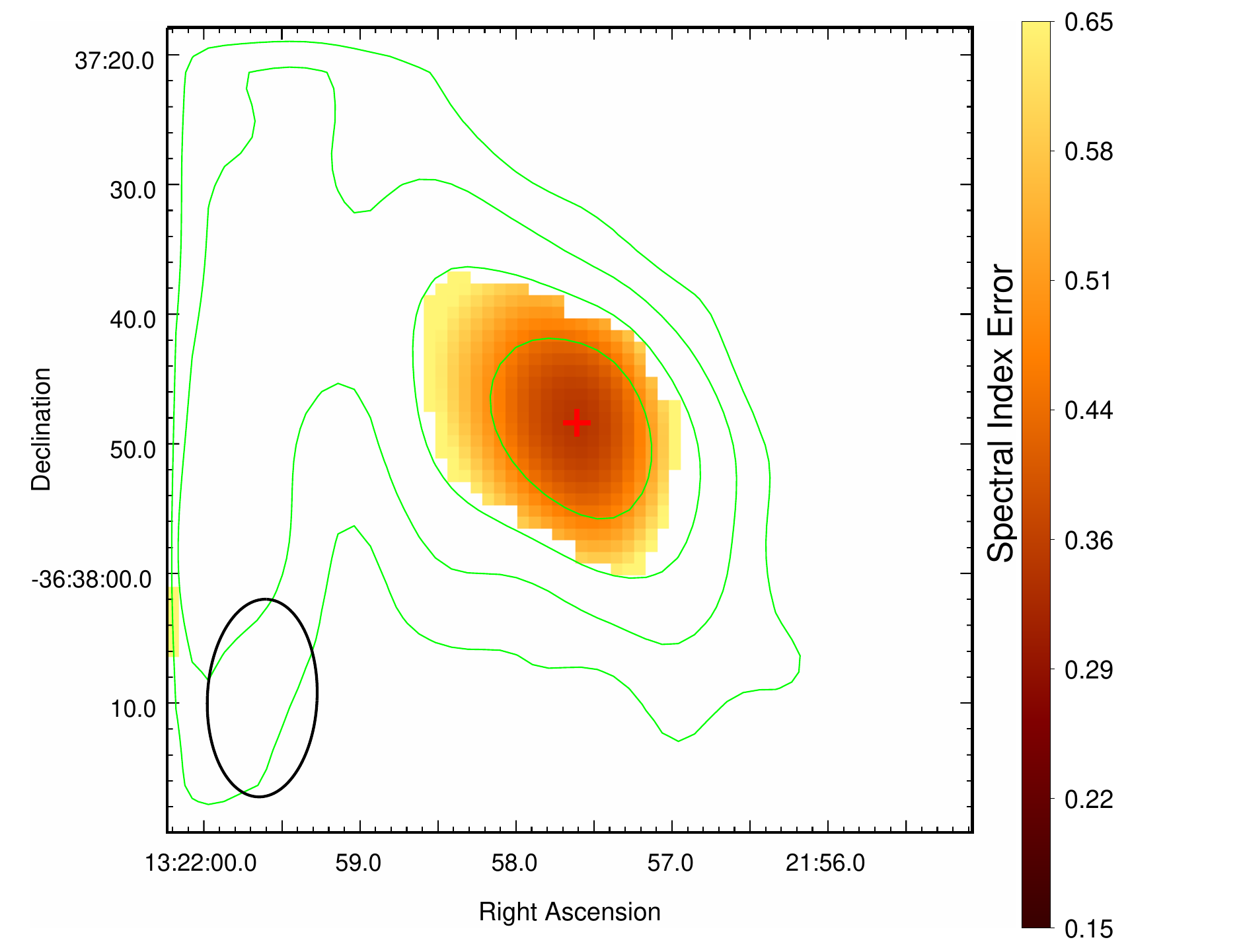}\\
    \caption{Left: Spectral index map for the ATCA/1.5D image, centered on the core of NGC\,5102. Right: Error on spectral index. Green contours taken from Figure \ref{fig:n5102_images}e for reference.}
    \label{fig:n5102_spid}
\end{figure*}

NGC\,205 was also observed by \textit{Chandra}/ACIS-S in 2004 (ObsID 4691, PI: Terashima). The observation has an exposure time of 8.9\,ks. No source was detected ($<3.4\times10^{-4}$ count\,s$^{-1}$).

\section{Results} \label{sec:results}

\subsection{NGC~5102}

Two X-ray point sources lie within $\sim5^{\prime\prime}$ of the nucleus of NGC\,5102 (Figure 1a). As stated above,
the dimmer, north-eastern source is coincident with the optical center of the nucleus of the galaxy to within 0.4\arcsec. Given this spatial coincidence, and the existence of a dynamically-confirmed SMBH at the center of the nucleus, the most straightforward explanation for the X-ray source is that it is a low-luminosity AGN. The source has an unabsorbed X-ray flux $S_{1-10\,\kev}= (3.0^{+1.5}_{-1.2})\times10^{-15}\flux$ and corresponding luminosity $L_{1-10\,\kev}= (5.0^{+2.5}_{-1.9})\times10^{36}\ergs$. However, we note that there is no clear optical evidence for an AGN (e.g., \citealt{2017MNRAS.464.4789M}), so we cannot entirely rule out the possibility that the X-ray source is an X-ray binary unrelated to the SMBH. Nevertheless, the strong evidence for radio emission associated with the SMBH (see below) implies recent accretion, which would support an AGN explanation for the nuclear X-ray source.

The second, brighter source is located $\approx3.5^{\prime\prime}$ to the south-west of the optical nucleus. This is inconsistent with the optical nucleus at the $>4\sigma$ level, so we can be confident that this source is not associated with the black hole, but instead is likely to be an unrelated X-ray binary. It has an unabsorbed flux of $S_{1-10\,\kev}= (2.1\pm0.2)\times10^{-14}\flux$ and luminosity $L_{1-10\,\kev}= (3.5\pm0.4)\times10^{37}\ergs$.

The radio emission in the ATCA image is clearly not a point source in both the extended 6\,km and confined 1.5\,km configurations. At both 1.6\,GHz and 2.6\,GHz, across all three observations, the extended emission appears centrally peaked, with the peak close to coincident with the optical nucleus (Figure \ref{fig:n5102_images}). In the 6A/B images, the integrated fluxes are all $\approx1$\,mJy and with the brightest emission spread over an area $\sim10^{\prime\prime}$ wide. Clearly, radio continuum emission is being resolved out in the higher-resolution images, as the low-resolution 1.5\,km configuration images detect a larger radio structure ($\sim20-40^{\prime\prime}$ in diameter). The integrated fluxes are also significantly larger: 9.3 and 3.4\,mJy for the 1.6 and 2.6\,GHz images, respectively. Based on all sets of ATCA data, we estimate that the brightest emission is between 20-80\,pc wide, though there are hints of faint emission on larger scales of hundreds of parsecs.

The ASKAP images have a a similar resolution to the ATCA 1.5\,km configuration images, but at a central frequency of $\sim 0.9$ GHz that is about a factor of two lower. In these data, taken about 6 years later than the ATCA data, the source has a peak flux of 1.2\,mJy beam$^{-1}$, but does not appear to be extended. Since the spatial scale of the emission is too large for a physical change to have occurred over this timescale, the difference between the ASKAP and low-resolution ATCA images is likely due to the different central frequency and higher noise level in the ASKAP data.

In the high-resolution C band VLA images, the extended source is not seen at all: it is likely that all of the emission is fully resolved out. In addition, there is no core emission seen at all at the location of the optical nucleus, to a $3\sigma$ upper limit of $< 14.3 \, \mu$Jy at an average frequency of 6 GHz. We convert this upper limit to a 5\,GHz luminosity, typical for use with the fundamental plane of accretion, assuming a flat spectral index; $L_{5\,GHz}<1.2\times10^{33}\, \ergs$.

\subsection{NGC~205}

\begin{figure*}[t]
    \centering
    \includegraphics[width=0.32\textwidth]{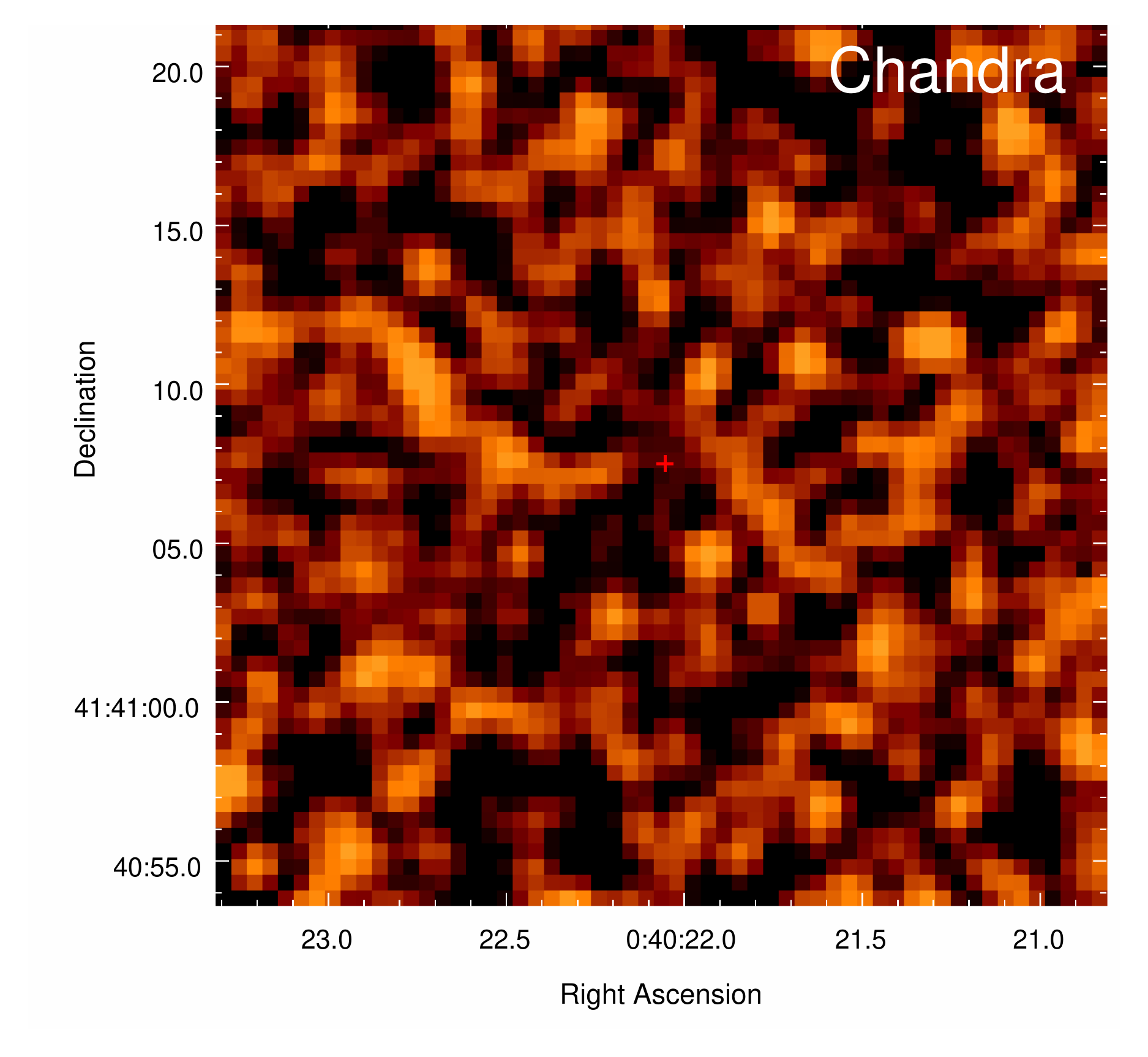}
    \includegraphics[width=0.32\textwidth]{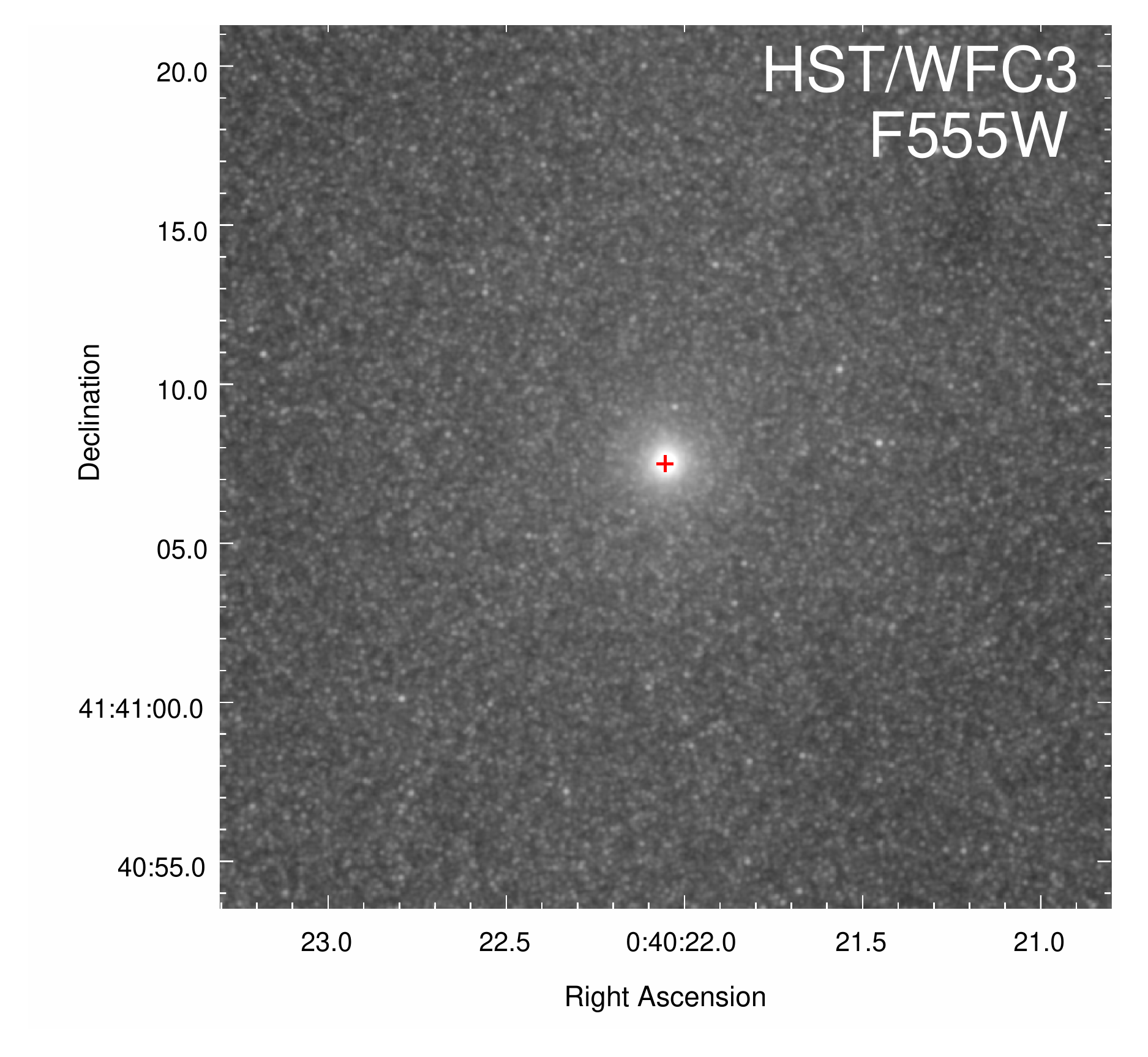}    
    \includegraphics[width=0.32\textwidth]{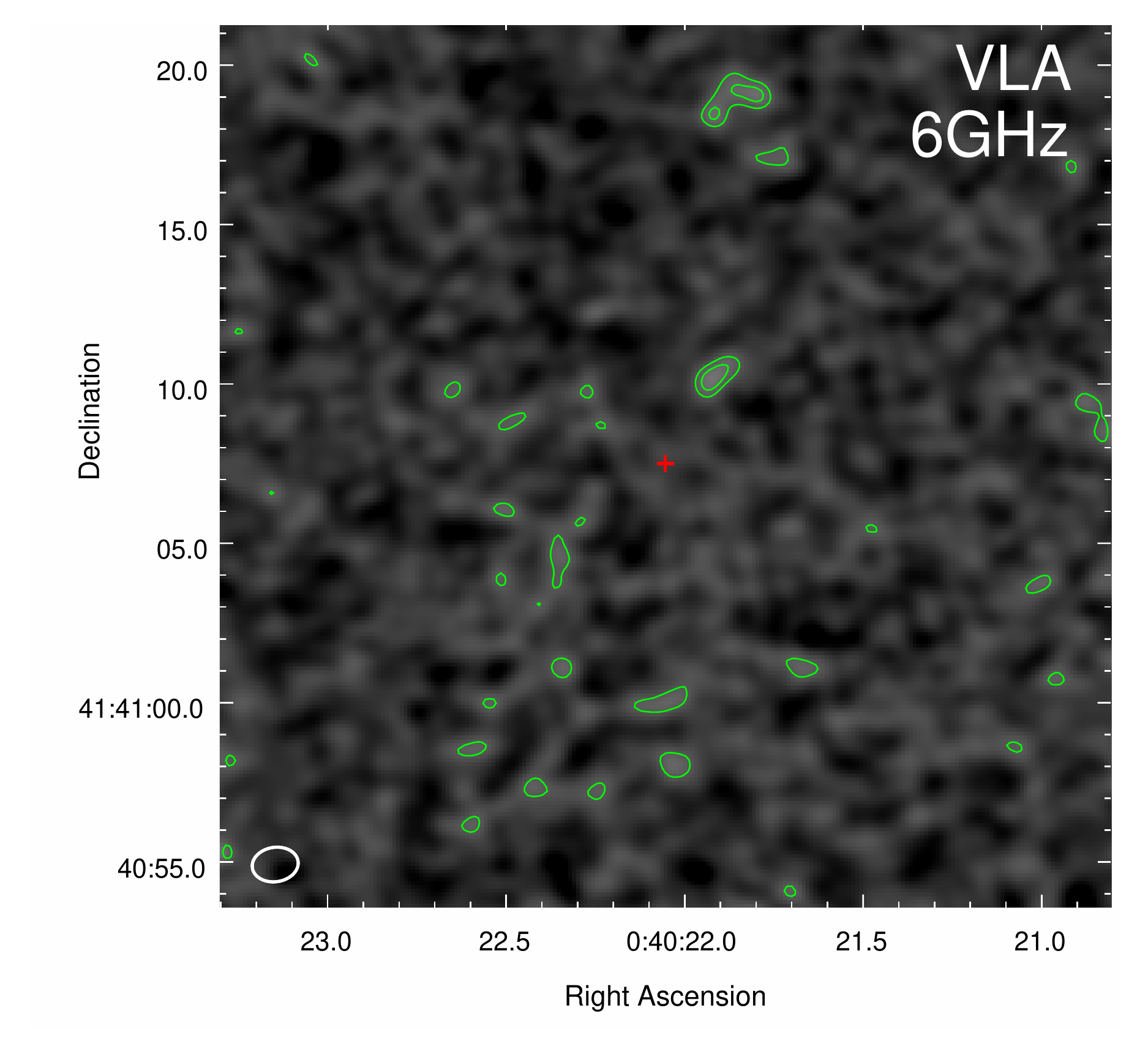}\\
    \caption{Left: \textit{Chandra} image of NGC\,205 smoothed using a Gaussian tophat 2 pixels wide for visualization purposes. The red cross marks the optical nucleus. Middle: As in left, but for the \textit{HST}/WFC3 F555W image. Right: As in left, but for the 6\,GHz VLA image. The beam size is indicated by the white ellipse in the lower left corner. Contours are set at $2\sqrt{2^n}\times \sigma $, where $\sigma$ is the local rms ($\sigma=1.1\,\mu$Jy beam$^{-1}$) and $n=1,2,3,4..$.  The source is not detected in either the radio or X-ray image.}
    \label{fig:n205_images}
\end{figure*}

No X-ray or radio source is detected at the position of the NGC\,205 optical nucleus (ICRS R.A.=00:40:22.054, Dec.=+41:41:07.50; Figure \ref{fig:n205_images}). We place an upper limit on the 1-10\,keV unabsorbed X-ray flux of $S_{1-10\,keV}<9.7\times10^{-16}$\,\flux and equivalent luminosity is $L_{1-10\,keV}<7.6\times10^{34}$\,\ergs. The 3$\sigma$ 6\,GHz radio upper limit is $S_{6\,GHz}<3.3\,$\fluxden, with a corresponding luminosity limit (again assuming a flat spectral index) of $L_{5\,GHz}<1.2\times10^{32}\ergs$.


\section{Discussion} \label{sec:discussion}

\subsection{Evidence for past AGN activity: NGC\,5102}

There is clear evidence of diffuse radio emission coincident with the nucleus of NGC\,5102. Based on our low-resolution 1.5D configuration ATCA observations, we find the size of the emission to be $\gtrsim 500\,$pc (at 2\,GHz). The spectral index map suggests that the average spectral index between $-1$ and $-1.5$, with steepening towards the east and west edges of the emission.

Here we discuss possible interpretations of this diffuse emission, concluding that it is mostly likely due to AGN activity that may have been higher in the past.


\subsubsection{Radio emission from gas heated by stars}

Extended radio structure could be a result of free-free emission from diffuse ionized gas within the nucleus of NGC\,5102. This explanation is strongly disfavored for two reasons: first, young hot stars would be required to ionize the interstellar madium (ISM), but there is no evidence of such a population within the nucleus of NGC\,5102 \citep{2018MNRAS.480.1973K, 2021AJ....162..281H}. Second, free-free emission tends to be flat-spectrum, while the integrated flux in NGC\,5102 appears steep, instead suggesting that the radio source is dominated by optically-thin synchrotron emission. We can estimate the free-free contribution based on the Balmer emission. \citet{2018MNRAS.480.1973K} measure a $H_{\beta}$ luminosity $L_{H_{\beta}}=1.9\times10^{36}\,\ergs$, suggesting a free-free flux density at 5\,GHz of only $3.5\,\mu$Jy (\citealt{1986A&A...155..297C}; Appendix A), negligible compared to our measured integrated flux. We note that in principle the Balmer emission may be underestimated due to line-of-sight extinction, though Figure 1(b) shows that there appears to be little or no dust on the nucleus itself. Overall, this explanation is highly inconsistent with the data.

\subsubsection{ULX radio bubble}

Ultraluminous X-ray sources (ULXs) are thought to be stellar-mass black hole or neutron star X-ray binaries accreting at super-Eddington rates \citep{2017ARA&A..55..303K}. They are primarily characterized by their extreme X-ray luminosities ($L_X \geq 10^{39}$\,\ergs). A handful of ULXs have shown evidence of extended emission which is thought to be a result of ultra-powerful outflows. Either through a highly collimated jet or fast disk wind, mechanical energy is injected into the surrounding ISM.  Over the lifetime of the ULX phase, an optical or radio ``ULX bubble'' is inflated. These ULX bubbles typically have radio luminosities $\sim10^{34-35}$ erg/s and are on the order of $\sim50-500$\,pc in diameter (e.g., \citealt{2010Natur.466..209P,2021MNRAS.501.1644S}). This matches our estimates on both the size and radio luminosity of the extended nuclear emission in NGC\,5102. 

Of course, the primary issue with this explanation is that the central X-ray source has only $L_{X}\sim 5 \times 10^{36}$ erg s$^{-1}$. Extreme obscuration could be one possibility, but even though a detailed X-ray spectral fit is not possible, this is ruled out by the detection of the source at $< 1.5$ keV \citep{2010ApJS..189...37E}.

Alternatively, the extended radio emission
in NGC\,5102 may have come from a ULX that has since ``turned off", becoming X-ray fainter by a factor of at least $\gtrsim 200$. This cannot be directly constrained by existing X-ray data, but is disfavored indirectly: ULXs typically reside in young stellar populations ($\lesssim10$\,Myr), while there is no evidence of such stars within the central region of NGC\,5102 \citep{2018MNRAS.480.1973K, 2021AJ....162..281H}.

\subsubsection{AGN outflows}

While there is currently no detected core radio emission in the nucleus of NGC\,5102, a previous phase of jet emission may be responsible for the extended nuclear radio emission. First, the emission is close to (but not exactly) symmetric around the optical nucleus, suggesting they are associated. In addition, from our lowest resolution ATCA observation, we find the radio structure to have a steep ($\alpha\lesssim-1$) spectral index (Figure \ref{fig:n5102_spid}). This is consistent with optically-thin synchrotron which is expected from
interactions between jet and surrounding ISM \citep{2001ApJ...559L..87N}.

The morphology of the radio emission varies among the radio images. In the lowest-resolution ATCA image (1.5D, 1.6\,GHz), it is clearly extended from the northeast to southwest, but the emission is faint, so it is unclear whether the morphology of this extension is due to the source itself or to background. In the higher-resolution 1.6 GHz images (both 6A and 6B configurations), it also appears to be extended along a similar, but not identical axis, with the southwest emission brighter. The emission in the different configurations at 2.6 GHz appears extended but with no particular preferred axis (Table \ref{tab:n5102_radio_shift}). In all the ATCA images, except for the 1.5D/2.6 GHz image, the peak of the radio emission is consistent with a small ($\sim 1.2\arcsec$ = $\sim 20-25$ pc) offset from the optical position. 

One interpretation of this offset is that it could be due to a jet pair partially angled to our line of slight, with the southern jet being the approaching, Doppler-boosted one, and hence slightly brighter. Another possibility is that the brighter emission south of the optical position does not reflect the Doppler-boosted jet power, but instead is due to another asymmetric physical cause, such as the distribution of diffuse gas the jet is encountering.

The integrated flux density at 1.6 GHz in the lowest-resolution ATCA image is $\sim 9.3$ mJy, which corresponds to a luminosity of $\sim 6 \times 10^{35}$ erg s$^{-1}$. This is likely a lower limit to the past total jet power, which also would include $PdV$ work done by the jet.

A similar radio structure is seen in the nucleus of the early-type dwarf galaxy M32, which lacks core radio emission but has extended emission on a spatial scale of $\sim 10$ pc, inferred to be synchrotron emission from AGN outflows interacting with the ISM \citep{2020ApJ...894...61P}. NGC\,404 also shows extended radio emission on a spatial scale $> 10$ pc, but unlike NGC\,5102 and M32, the radio continuum emission is centrally peaked, consistent with core emission and hence current jet activity \citep{2017ApJ...845...50N}.

Given the lack of other likely explanations, and the partial similarities to other nearby low-luminosity AGN, the extended radio emission in NGC\,5102 is best explained as a radio lobe inflated by the AGN jet during a previous phase of activity that has since become substantially weaker or stopped entirely. The main difference in NGC\,5102 is that in addition to emission on spatial scales of tens of parsecs, there is also fainter, more diffuse emission visible. 
 

\subsection{Constraints on current AGN activity: NGC\,5102 and NGC\,205}

\subsubsection{NGC\,5102}

The 2002 Chandra detection of a $\sim 5\times10^{36}$ erg s$^{-1}$ X-ray source at the center of NGC\,5102, consistent with the position of its dynamically detected SMBH, is consistent with a low-luminosity AGN having been active at that time. Using the fundamental plane of black hole activity \citep{2019ApJ...871...80G} predicts that \emph{if} the X-ray luminosity at the time of the 2021 VLA 6 GHz observations were the same as in 2002, then the predicted flux density would have been $\sim 43 \mu$Jy. This is about a factor of 3 brighter than the $3\sigma$ upper limit from the VLA data, but within the scatter of the fundamental plane. This comparisons suggests that if NGC\,5102 falls on the fundamental plane, it is unlikely that the source was substantially more luminous in the X-rays in 2021 compared to 2002, though it could be fainter.  We emphasize that for this comparison we are only using the fundamental plane as a tool: since there are no deep simultaneous X-ray and radio measurements for NGC\,5102, it provides no new constraints on the fundamental plane.

Since we do not have broadband spectral information for NGC\,5102, the X-ray luminosity is the main constraint we have on its bolometric luminosity. We use a standard bolometric correction factor of $L_{\rm bol}/L_{2-10\,keV}=15.8$ \citep{2008ARA&A..46..475H} for the 2--10 keV X-ray band, which assumes that AGN in low-mass galaxies have similar spectral energy distributions to more massive galaxies.

For  NGC\,5102, $L_{\rm bol} = 5.2\times10^{37}\,$\ergs, which corresponds to $L_{\rm bol}/L_{\rm Edd} = 4.4\times10^{-7}$, if the X-ray emission is associated with an AGN. If the X-ray emission is instead due to an X-ray binary, then this is an upper limit on $L_{\rm bol}/L_{\rm Edd}$. While this is our best estimate of the Eddington ratio in 2002, the existence of extended radio emission (and the corresponding lack of present core radio emission) suggests $L_{\rm bol}/L_{\rm Edd}$ was higher at some point in the past. 

\subsubsection{NGC\,205}

There is no evidence for an AGN in NGC\,205, for which we have stringent upper limits on its X-ray and radio emission. We convert the upper limit on its X-ray luminosity to a limit on its bolometric luminosity using the same standard correction factor, giving $L_{\rm bol} < 4.8\times10^{35}\,$\ergs\ from our 2020 \textit{Chandra} data. This is equivalent
to $L_{\rm bol}/L_{\rm Edd} < 5.6\times10^{-7}$ at the median dynamical mass of $6800\,M_{\odot}$, but at the upper dynamical mass limit of $\sim 10^5 M_{\odot}$, the value would be much lower, with $L_{\rm bol}/L_{\rm Edd} < 3.7\times10^{-8}$.






\subsection{Context: Building Local and Comparison Samples}

\begin{figure*}[t]
    \centering
    \includegraphics{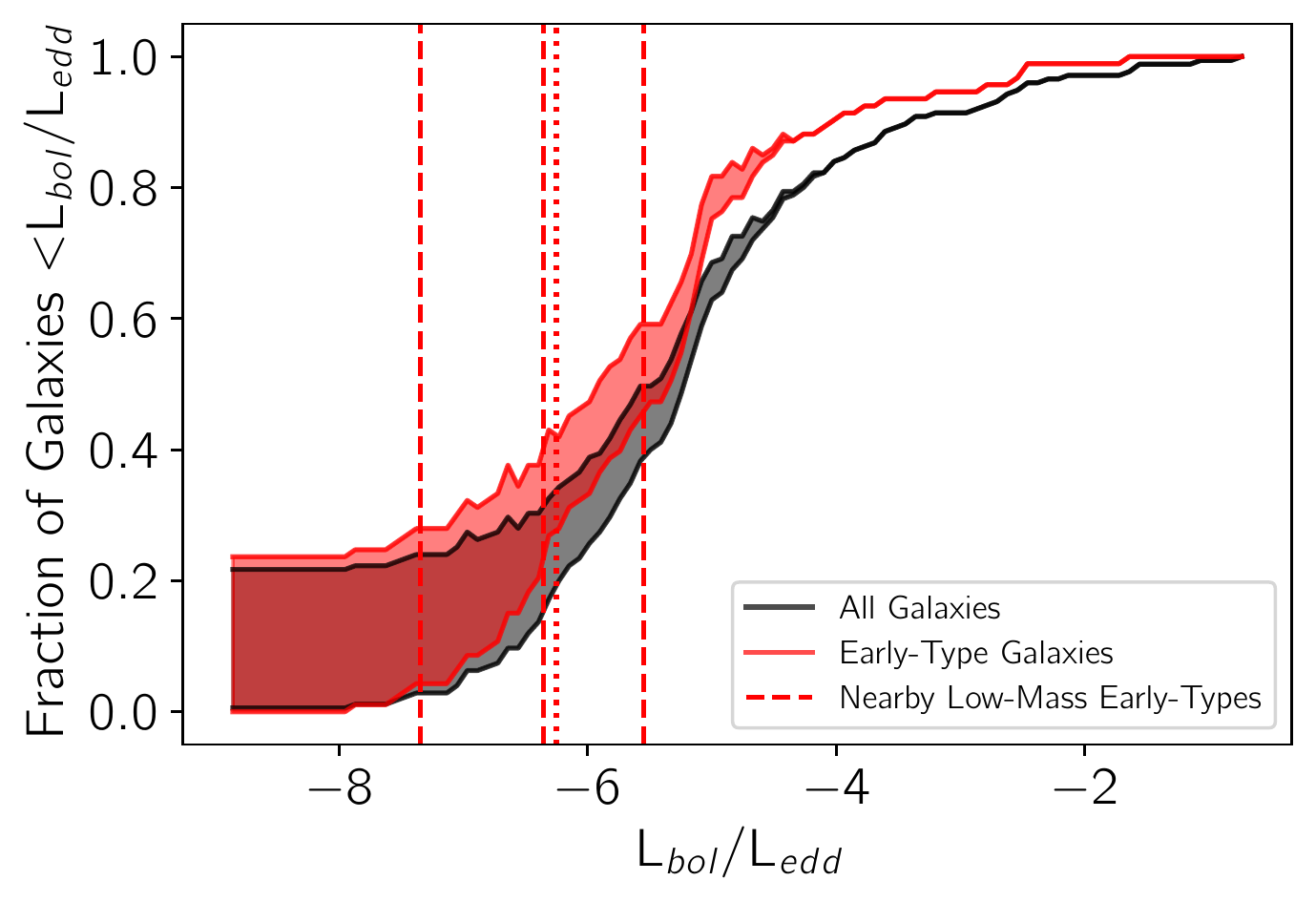}
    \caption{The nearest low-mass early-type galaxies have low accretion rates.  The solid lines and regions show the cumulative distribution function of $L_{\rm bol}/L_{\rm Edd}$ distributions of bright, massive galaxies based on X-ray measurements from \citet{2009ApJ...699..626H} assuming $L_{\rm bol} = 15.8 L_{X}$. The black lines shows the full galaxy sample (175 galaxies), while the red lines show just the early-type galaxies (93 galaxies). The shaded regions in both samples show the impact of upper limits on the cumulative distribution functions. The three dashed lines, from left to right, show the $L_{\rm bol}/L_{\rm Edd}$ measurements for M32, NGC\,5102, and NGC\,404, while the dotted line shows the upper limit on $L_{\rm bol}/L_{\rm Edd}$ for NGC\,205, assuming a black hole mass of 6800 M$_\odot$.}
    \label{fig:lbol}
\end{figure*}

Central black holes have also been detected in the nearby low-mass early-type galaxies M32 \citep{1998ApJ...493..613V,2018ApJ...858..118N}, NGC\,404 \citep{2010ApJ...714..713S,2017ApJ...836..237N,2020MNRAS.496.4061D}, and NGC\,5206 \citep{2018ApJ...858..118N,2019ApJ...872..104N}. Together with NGC 5102 and NGC 205, this is a volume-limited sample of nearby $10^{9}$ --$10^{10} M_{\odot}$ early-type galaxies.

While NGC 5206 has no \emph{Chandra} or even \emph{XMM} data, we can use the luminosity of the central X-ray source in both NGC 404 \citep{2011ApJ...737...77B,2012ApJ...753..103N} and M32 \citep{2003ApJ...589..783H} to estimate their bolometric Eddington ratios in the same manner as for NGC 5102 and NGC 205, finding values of $L_{\rm bol}/L_{\rm Edd} = 2.8\times10^{-6}$ (NGC 404) and 
$L_{\rm bol}/L_{\rm Edd} = 4.5\times10^{-8}$ (M32). We note that a few low-luminosity AGN (including that in M32) appear to be quite bright in the infrared. This emission could contribute to or even dominate their bolometric luminosity 
\citep{2010ApJ...725..670S,2010ApJ...714..713S,2020ApJ...888...19D}, but here we use a fixed bolometric conversion factor to enable a consistent comparison to other studies.


In Figure \ref{fig:lbol}, we can see that all four of these nearby low-mass early-type galaxies are at or below the median $L_{\rm bol}/L_{\rm Edd}$ of massive galaxies from \citet{2009ApJ...699..626H}. This Eddington ratio distribution is based on 175 galaxies with archival \textit{Chandra} X-ray measurements out of the 486 galaxy Palomar galaxy sample \citep{1997ApJS..112..315H}. We note that there are two selection criteria that may bias the distribution in \citet{2009ApJ...699..626H} to higher $L_{\rm bol}/L_{\rm Edd}$ values than a volume-limited sample: (i) all galaxies with nearby \textit{Chandra} archival data are used, including data that may have been taken due to the presence of a known AGN, which would tend to select for higher Eddington ratios, and (ii) all galaxies with optical emission lines classified as star-forming galaxies are excluded; these galaxies will likely have typically lower $L_{\rm bol}/L_{\rm Edd}$ compared to those with optical emission classified as AGN. To mitigate this latter effect, in Figure \ref{fig:lbol} we also plot just galaxies with Hubble Type $T <= 0$ (early-type galaxies, including some S0/Sa transition objects).
The potential bias due to the exclusion of star-forming galaxies has a minimal impact on this subsample. 


The main result is that the four nearby low-mass early-type galaxies have $L_{\rm bol}/L_{\rm Edd}$ at or below the median of the distribution of typically more massive early-type galaxies from \citet{2009ApJ...699..626H}. NGC\,404 is close to the median measured $L_{\rm bol}/L_{\rm Edd}$ of $\sim 5 \times 10^{-6}$, while the other three galaxies are below, at somewhat lower Eddington ratios. 

To compare the distributions statistically, we first note that both the comparison sample and the nearby low-mass galaxy sample have a meaningful fraction of measurements that are upper limits, which challenges the use of typical non-parametric tests or a simple bootstrap. Since the observed $L_{\rm bol}/L_{\rm Edd}$ values for early-type galaxies appear relatively well-represented by a lognormal distribution, we fit a lognormal model to the data using the Bayesian Markov Chain Monte Carlo software {\tt JAGS} \citep{Plummer2012}, self-consistently incorporating the upper limits. The resulting best-fit values are $\mu = -5.69\pm0.15$ and $\sigma = 1.36\pm0.12$. This mean corresponds to $L_{\rm bol}/L_{\rm Edd} = 2 \times 10^{-6}$, just below the measured value for NGC\,404. When we fit a lognormal model to the low-mass nearby galaxy sample, again modeling the upper limit (for NGC\,205) self-consistently, we find $\mu =  -6.57\pm0.50$ and $\sigma = 0.96\pm0.44$. The difference in means is $-0.88\pm0.52$, corresponding to $p=0.046$. Hence, with this current small sample the evidence for a difference in mean $L_{\rm bol}/L_{\rm Edd}$ between low and high-mass galaxies is only suggestive and is far from conclusive. Another interpretation of this result, especially given the very low accretion rates in the Milky Way \citep{1998ApJ...492..554N} and M31 \citep{2011ApJ...728L..10L} is that there is a bias towards higher Eddington ratios in the \citet{2009ApJ...699..626H} sample, perhaps due to X-ray binary contamination of nuclear sources.



\subsection{Interpretation and Future Surveys}

The suggestive, albeit preliminary evidence that lower-mass galaxies could have lower Eddington ratios than higher-mass galaxies is in mild tension with existing literature on the X-ray luminosities of galaxy nuclei as a function of stellar mass.  In particular, both \citet{2015ApJ...799...98M} and \citet{2019ApJ...883L..18G} model the X-ray luminosities of a large number of early-type galaxies with a model that includes a varying black hole occupation fraction and a distribution of X-ray luminosities that depends on the galaxy stellar mass.  Both papers find the slope of the $L_X$--$M_\star$ relation is $\lesssim 1$. These $L_X$--$M_\star$ relations are fit using much shallower data than we present here and are mostly upper limits at $M_\star < 10^{10} M_{\odot}$. Thus the $L_X$ distribution of our deeper observation can provide a useful check on their assumptions of a $L_X$--$M_{\star}$ relation with constant scatter and a slope $\lesssim 1$.

We find that all four of the nearest early-type galaxies fall below the mean of the best-fit $L_X$--$M_{\star}$ relation from \citet{2015ApJ...799...98M}, with each of the M\,32 and NGC\,5102 measurements and the NGC\,205 upper limit all being quite low outliers. If the low X-ray luminosities tentatively seen in our current small sample is borne out in larger samples, it would suggest that the $L_X$--$M_\star$ relation is either steeper or has an increased scatter at $M_\star < 10^{10} M_{\odot}$. A steeper $L_X$--$M_{\star}$ relation could result from some combination of (i) lower Eddington fractions at lower $M_{\star}$, as suggested in the present paper; (ii) a steeper $M_{BH}$--$M_{\star}$ relation (e.g., \citealt{2019ApJ...883L..18G}), or (iii) potentially from a systemic change in the spectral energy distributions of AGN in lower-mass galaxies, such that a much smaller fraction of the bolometric luminosity is emitted in X-rays.



Another potential implication is in using multiwavelength data to constrain the occupation fraction of central black holes: at the inferred mean $L_{\rm bol}/L_{\rm Edd}$ of our sample ($3 \times 10^{-7}$), a $5 \times 10^5 M_{\odot}$ black hole would have $L_X \sim 10^{36}$ erg s$^{-1}$, which is far too low to confidently attribute to an AGN without other evidence such as radio emission. It also suggests generically that any survey that assumes an Eddington ratio distribution typical of more massive galaxies could potentially underestimate the black hole occupation fraction \citep[e.g.,][]{2015ApJ...799...98M}. Our results also show that it is challenging to test the black hole fundamental plane for low-mass central black holes, which inhibits its utility in reaching broad conclusions about the occupation fraction and mass distribution of central black holes in low-mass galaxies.

These results could also be strengthened by obtaining high-quality spectral energy distributions for a subset of the nearby dynamically confirmed low-mass central black holes. Such work is feasible with $JWST$, and should be an important priority moving forward. A comprehensive survey for extended radio continuum emission in nearby low-mass galaxies, possible even with present telescopes, could also reveal past AGN emission in galaxies that are quiescent today.

\section{Conclusions}

We have analyzed X-ray and radio observations for two nearby low-mass galaxies with dynamically confirmed central black holes, NGC\,5102 and NGC\,205. In NGC\,5102, we find mixed evidence for current AGN activity, with a nuclear X-ray source but no core radio continuum emission, but strong evidence for past AGN activity, with extended luminous radio continuum emission centered on the nucleus.

For NGC\,205, our simultaneous X-ray and radio observations have produced a state-of-the-art non-detection, placing strong constraints on the current accretion activity of the black hole.

We assess these results in the context of the other nearly low-mass early-type galaxies with dynamically confirmed central black holes, finding a hint that low-mass early-type galaxies have unusually low values of $L_{\rm bol}/L_{\rm Edd}$, though this conclusion is dependent both on a small sample of low-mass galaxies and potential biases in the comparison sample of more massive galaxies. These preliminary inferences are consistent with a scenario where the occupation fraction of central black holes in low-mass galaxies is high, but most of these sources will be difficult to confirm without dynamical measurements.

\begin{acknowledgements}

We thank Jenny Greene for a thoughtful reading of the paper and Tracy Clarke for useful discussions. We acknowledge the comments of an anonymous referee, which helped improve the presentation and context of our results.

We acknowledge support from NASA grant Chandra-GO0-21096X and the Packard Foundation.

\end{acknowledgements}

\software{ciao (v4.11; \citealt{2006SPIE.6270E..1VF}), DS9 (v8.1; \citealt{2003ASPC..295..489J}), CASA (v5.4.1; \citealt{2007ASPC..376..127M}), JAGS (v5.3.1; \citealt{Plummer2012})}

\bibliography{references}{}
\bibliographystyle{aasjournal}



\begin{table*}
    \centering
    \caption{Radio Observations of NGC\,5102}
    \begin{tabular}{cllccrrrr}
        \hline
        Fig.\ 1 & \multicolumn{1}{c}{Observation} & \multicolumn{1}{c}{Telescope} & \multicolumn{1}{c}{$\nu$} & \multicolumn{1}{c}{Beam} & \multicolumn{1}{c}{rms} & \multicolumn{1}{c}{$S_{peak}$} & \multicolumn{1}{c}{$S_{int}$} & \multicolumn{1}{c}{Source Size} \\
        Panel & Date& & (GHz) & \multicolumn{1}{c}{($^{\prime\prime}\times^{\prime\prime}$)} & \multicolumn{1}{c}{(mJy/beam)} & \multicolumn{1}{c}{(mJy/beam)} & \multicolumn{1}{c}{(mJy)} & \multicolumn{1}{c}{($^{\prime\prime}\times^{\prime\prime}$)}\\\hline\hline
        (c) & 2012-12-13  & ATCA/6B & 1.6 & $14.1\times6.4$ & 0.05 & $0.43\pm0.05$ & $1.01\pm0.08$ &  $12.9\times5.9$ \\
        (d) & 2012-12-13  & ATCA/6B & 2.6 & $10.0\times4.6$ & 0.03 & $0.25\pm0.03$ & $1.02\pm0.05$ & $12.2\times11.0$ \\
        (e) & 2012-12-20  & ATCA/1.5D & 1.6 & $15.3\times8.5$ & 0.17 & $0.93\pm0.17$ & $9.3\pm0.5$ &  $44.6\times27.2$ \\
        (f) & 2012-12-20  & ATCA/1.5D & 2.6 & $10.2\times5.5$ & 0.04 & $0.29\pm0.04$ & $3.4\pm0.14 $ & $31.0\times19.6$ \\
        (g) & 2013-02-11  & ATCA/6A & 1.6 & $13.3\times5.9$ & 0.03 & $0.38\pm0.03$ & $1.16\pm0.06$ & $15.7\times9.5$ \\
        (h) & 2013-02-11  & ATCA/6A & 2.6 & $8.8\times4.1$ & 0.02 & $0.18\pm0.02$ & $0.89\pm0.04$&  $17.0\times8.3$ \\
        (i) & 2019-04-29 & ASKAP & 0.9 & $15.9\times12.7$ & 0.2 & $1.2\pm0.2$ & - & - \\
        \hspace{2mm}(k)$^{\alpha}$ & 2021-01-27 & VLA & 4.9 & $1.28\times0.38$ & $0.0069$ & $<0.021$ & - &  - \\
         & 2021-01-27 & VLA & 7.0 & $0.92\times0.27$ & $0.0066$ & $<0.020$ & - &  - \\\hline
    \end{tabular}
    \begin{flushleft}
    \qquad$^\alpha$ Panel (k) in Figure \ref{fig:n5102_images} is the full band image (4-8\,GHz) centered at 6\,GHz.
\end{flushleft}
    \label{tab:n5102_radio_obs}
\end{table*}

\newpage

\begin{table*}
    \centering
    \large
    \caption{X-ray/radio luminosities}
    \begin{tabular}{lrrrr}
        \hline
        \multicolumn{1}{c}{Source} & \multicolumn{1}{c}{Mass} & \multicolumn{1}{c}{Distance} & \multicolumn{1}{c}{$L_{5GHz}$ $^a$} &  \multicolumn{1}{c}{$L_{1-10\,keV}$} \\
         & \multicolumn{1}{c}{($M_{\odot}$)} & \multicolumn{1}{c}{(Mpc)} & \multicolumn{1}{c}{(erg s$^{-1}$)} &  \multicolumn{1}{c}{(erg s$^{-1}$)} \\\hline\hline 
         NGC\,5102 & $9.12^{+1.84}_{-1.53}\times10^{5}$ & 3.74 & $<1.8\times10^{33}$ & $4.4^{+2.2}_{-1.7}\times10^{36}$ \\
         NGC\,205 & $6.8^{+95.6}_{-6.7}\times10^{3}$ & 0.81 & $<1.2\times10^{32}$ & $<3.4\times10^{34}$ \\\hline
    \end{tabular}
\begin{flushleft}
    \qquad\qquad\qquad\qquad\quad $^\alpha$ Assumes a flat spectral index.
\end{flushleft}
    \label{tab:lrlx}
\end{table*}

\begin{table*}
    \centering
    \caption{NGC\,5102 radio emission offset}
    \begin{tabular}{cllcccr}
        \hline
        Fig.\ 1 & \multicolumn{1}{c}{Observation} & \multicolumn{1}{c}{Telescope} & \multicolumn{1}{c}{$\nu$} & \multicolumn{2}{c}{Offset$^\alpha$} & \multicolumn{1}{c}{Direction} \\
        Panel & Date& & (GHz) & \multicolumn{1}{c}{RA($^{\prime\prime}$)} & \multicolumn{1}{c}{Dec($^{\prime\prime}$)} & \\\hline\hline
        (c) & 2012-12-13  & ATCA/6B & 1.6 & 0.1 & 1.0 & South-West \\
        (d) & 2012-12-13  & ATCA/6B & 2.6 &  0.5 & 1.3 & South-West \\
        (e) & 2012-12-20  & ATCA/1.5D & 1.6 & -0.3 & 0.4 & South-East \\
        (f) & 2012-12-20  & ATCA/1.5D & 2.6 & -0.4 & -0.7 & North-East \\
        (g) & 2013-02-11  & ATCA/6A & 1.6 & 0.4 & 1.6 & South-West\\
        (h) & 2013-02-11  & ATCA/6A & 2.6 & -0.5 & 1.2 & South-East \\\hline
    \end{tabular}
    \begin{flushleft}
    \qquad\qquad\qquad\qquad$^\alpha$ Shift from the Gaia position of the nucleus of NGC\,5102 to the flux peak of the extended radio emission.
\end{flushleft}
    \label{tab:n5102_radio_shift}
\end{table*}

\end{document}